\documentclass[aps,twocolumn,longbibliography,floatfix]{revtex4-1}
\usepackage{graphicx}
\usepackage{bm}
\usepackage{color}
\usepackage{amsmath}
\usepackage{mathtools}
\usepackage{cancel}
\usepackage[super]{nth}

\begin{document}
\title{Effect of Non-Heisenberg Magnetic Interactions on Defects in Ferromagnetic Iron}
\author{Jacob B. J. Chapman}
\email{Corresponding author: jacob.chapman@ukaea.uk}
\affiliation{UK Atomic Energy Authority, Culham Science Centre, Oxfordshire, OX14 3DB, United Kingdom}
\author{Pui-Wai Ma}
\affiliation{UK Atomic Energy Authority, Culham Science Centre, Oxfordshire, OX14 3DB, United Kingdom}
\author{Sergei L. Dudarev}
\affiliation{UK Atomic Energy Authority, Culham Science Centre, Oxfordshire, OX14 3DB, United Kingdom}

\begin{abstract}

Fundamental flaws in the Heisenberg Hamiltonian are highlighted in the context of its application to BCC Fe, including the particular issues arising when modelling lattice defects. Exchange integrals are evaluated using the magnetic force theorem. The bilinear exchange coupling constants are calculated for all the interacting pairs of atomic magnetic moments in large simulation cells containing defects, enabling a direct mapping of the magnetic energy onto the Heisenberg Hamiltonian and revealing its limitations. We provide a simple procedure for extracting the Landau parameters from DFT calculations, to construct a Heisenberg-Landau Hamiltonian. We quantitatively show how the Landau terms correct the exchange-energy hypersurface, which is essential for the accurate evaluation of energies and migration barriers of defects.

\end{abstract}
\maketitle


\section{Introduction}


Magnetism is a quantum mechanical phenomenon that arises from a combination of the Coulomb interaction between electrons and the Pauli exclusion principle. The spin state of the electrons affects the total energy through what is known as exchange interaction. In transition, rare earth, and actinide  metals \cite{Hay_jacs_1975}, electrides \cite{kim_prb_2018} and organic polyradicals \cite{Lineberger_pccp_2011}, which all have partially filled $d$ or $f$-orbitals, magnetic moments are formed due to the exchange interaction between intra-atomic $d$ or $f$-electrons. Magnetism has been highly influential on modern technologies such as magnetic storage \cite{Jiles1998} and spintronic devices \cite{Shinjo2009}. Exotic non-collinear spin-textures such as skyrmions promise to revolutionise processor and data storage technologies further \cite{Seki2016}.

Iron-based alloys are particularly important industrial materials. They attain a myriad of complex magnetic states, such as ferro- and antiferromagnetic \cite{Pepperhoff2010}, incommensurate spin density waves \cite{Overhauser_PR_1962, Fawcett_RMP_1988, Burke_JPF_1983} and spin-glasses \cite{Burke_JPF_1983_b,Chapman_prb_2019}. Their mechanical properties are partially governed by the population of magnetic states \cite{Waggoner_1912,Mergia_2008}. For example, in pure iron, the softening of the tetragonal shear modulus $C'$ near the Curie temperature $T_C$ is driven by magnetism \cite{Hasegawa_JPhysF_1985,Dever_JAP_1972,Razumovskiy2011}.

Body-centred cubic (BCC) crystal structure of iron owes its stability to the free energy contributions from both lattice and magnetic excitations \cite{Hasegawa_PRL_1983,Kormann2008,Lavrentiev_PRB_2010,Lavrentiev_PRB_2011,Kormann2016,ma_prb_2017}. Magnetism also makes the $\langle 110 \rangle$ dumbbell the most stable configuration of a self-interstitial atoms (SIA) in iron. This is in contrast to other non-magnetic BCC transition and simple metals where a single SIA defect adopts a $\langle 111 \rangle$ or $\langle 11\chi\rangle$ configuration  \cite{nguyen-manh_prb_2006,derlet_prb_2007,Ma_PRM_2019}. 

The Heisenberg Hamiltonian \cite{blundell2001} is a well known model describing interaction between magnetic moments. It assumes that electrons are reasonably well localised, which is indeed the case in metals with $d$ or $f$-electrons. The Heisenberg Hamiltonian can be written as:
\begin{equation}\label{eqn:hhe}
    \hat{\mathcal{H}} =- \displaystyle \sum _{i \neq j} J_{ij}^{\text{eff}} \mathbf{\hat{s}}_i\cdot \mathbf{\hat{s}}_j
\end{equation}
where $\mathbf{\hat{s}}_i$ is a unit vector in the direction of an atomic spin $\mathbf{S}_i = S_i \mathbf{\hat{s}}_i$ at site $i$. $J_{ij}^{\text{eff}}$ is an effective isotropic pairwise exchange coupling parameter describing interaction between spins at sites $i$ and $j$. The local atomic magnetic moment and spin at site $i$ are related simply by $\mathbf{M}_i=-g\mu_B \mathbf{S}_i$ where $g=2.0023$ is the electron g-factor and $\mu_B$ is the Bohr magneton.

In the Heisenberg approximation, parameters $J_{ij}^{\text{eff}}$ govern the magnetic order, transition temperature and magnon dispersion of the material \cite{LKAG_jmmm_1987,Steenbock_jctc_2015,Kvashnin_prl_2016,Szilva_prl_2013,Szilva_prb_2017,Korotin_prb_2015,cardias_scirep_2017}. The value of $J_{ij}^{\text{eff}}$ can be estimated from experimental observations by fitting the temperature-dependent magnetic susceptibility curve \cite{trtica_inorgchem_2010,Abedi_dalton_2011}. On the other hand, $J_{ij}^{\text{eff}}$ can be determined from density functional theory (DFT) calculations \cite{andersen_prl_1984,LKAG_jmmm_1987,Schilfgaarde_jap_1999}. 


There are two commonly used approaches to deriving $J_{ij}^{\text{eff}}$ from DFT calculations. The first is the real-space total energy method \cite{Xie_PRB_2017}. The total energy is evaluated for various metastable collinear magnetic configurations. The exchange coupling parameter is then estimated from the energy differences between the various magnetic states.

This approach has several limitations. The necessity to perform total energy calculations for multiple configurations can be expensive in the limit of a large system size. This size problem cannot be circumvented for classes of materials such as organics or electrides \cite{kim_prb_2018}. In addition, the assumption that $J_{ij}^{\text{eff}}$ is a simple scalar does not help deliver information about the contributing orbitals or the dominant mechanism of exchange interaction.

The second approach is known as the Magnetic Force Theorem (MFT) \cite{LKAG_jmmm_1987,Lichtenstein_jpfmp_1984,Oguchi_jphysf_1983,Bruno_prl_2003,Steenbock_jctc_2015,Kvashnin_prl_2016,Szilva_prl_2013,Szilva_prb_2017,Yoon_PRB_2018,Korotin_prb_2015,cardias_scirep_2017,Han_prb_2004}. It was first derived for Ruderman-Kittel-Kasuya-Yoshida (RKKY) interactions between impurities in metals \cite{Lichtenstein_jpfmp_1984} using multiple scattering theory. The seminal idea led to the Lichtenstein-Katsnelson-Antropov-Gubanov (LKAG) equation \cite{LKAG_jmmm_1987}. This Green's function based approach provides an analytical expression for parameter $J_{ij}^{\text{eff}}$ in the form of a response to the changes in the total energy resulting from small spin rotations in a particular magnetic state.  

The principal advantage of the MFT approach is that all the pairwise parameters $J_{ij}^{\text{eff}}$ can be determined for a single magnetic configuration. The configuration does not need to be the true magnetic ground state, which can remain unknown. In addition, $J_{ij}^{\text{eff}}$ may be decomposed into contributions from different orbitals \cite{Yoon_PRB_2018,Korotin_prb_2015,cardias_scirep_2017}. 

Early developments of MFT were implemented using the localised orbital methods such as the linear muffin tin orbital (LMTO) approach \cite{Schilfgaarde_jap_1999,Katnelson_prb_2000} and for the linear combinations of pseudo-atomic orbitals (LCPAO)\cite{Han_prb_2004,Yoon_PRB_2018}. Recent extensions to plane-wave DFT codes have taken advantage of maximally localised Wannier functions \cite{Korotin_prb_2015}. 


Calculations of $J_{ij}^{\text{eff}}$ are often motivated by the need to parameterise multiscale methods as the Heisenberg model approach can then be used to predict finite temperature properties of magnetic systems \cite{ma_prb_2017,evans_jpcm_2014,Tranchida_jcompphys_2018, ma_pre_2010,ma_prb_2017,Lavrentiev_CompMatSci_2010,Lavrentiev_PRB_2010}. 
The studies performed using the MFT primarily concerned bulk materials or molecular magnets \cite{LKAG_jmmm_1987, Steenbock_jctc_2015,Boukhvalov_prb_2002,Boukhvalov_prb_2004}. Defects \cite{Boukhalov_prl_2007,Chang_prb_2007} as well as nanostructures on substrates \cite{Cardias_prb_2016} have also been considered. Nonetheless, even for perfect crystalline configurations it has been observed that the adiabatic magnetic exchange-energy hypersurface parameterised by the bilinear Heisenberg Hamiltonian is incomplete \cite{Drautz2005,Okatov_prb_2011,Singer_prl_2011,Singer2011a}. An accurate representation necessitates longitudinal fluctuations to be considered \cite{Singer_prl_2011,Ruban_prb_2007,Ma_prb_2012}.

Despite the known shortcomings of the Heisenberg Hamiltonian, it remains a popular choice for multiscale modelling. In this paper we address the consequences of the Heisenberg functional form of a magnetic Hamiltonian and show that applications of this Hamiltonian to distorted lattice configurations require extending it to the Heisenberg-Landau form \cite{Lavrentiev_PRB_2010,Lavrentiev_PRB_2011,Derlet2012}. We begin by benchmarking our density functional theory (DFT) calculations, performed using the OpenMX code \cite{openmx}, against the known literature and our in-house exchange coupling codes in Section~\ref{ss:bulk}. We then quantify the error of an idealised mapping of the DFT magnetic energy for pristine and defected configurations of body-centred cubic $\alpha$-Fe in Section~\ref{ss:sia}. 

Our analysis in Section~\ref{ss:fail} reveals that the magnetic hypersurface of point defects can be represented qualitatively, but the Heisenberg approximation fails to capture the relative stability of the $\langle 111 \rangle$ crowdion due to the mixed $e_g$-$t_{2g}$ characteristic of the bands at the Fermi energy. We assert that even a perfectly mapped Heisenberg Hamiltonian is unable to predict point defect behaviour in iron with reasonable quantitative reliability. In Section~\ref{ss:landau} we demonstrate that a very accurate representation can be created by incorporating in the Hamiltonian the \nth{2} and \nth{4} order Landau coefficients. We provide a simple procedure showing how to extract atom-resolved Landau parameters from DFT calculations. This enables the itinerant behaviour of $d$ electrons to be incorporated into the magnetic model. Finally, in Section~\ref{sec:mig} we explore the effect of magnetic interactions on the migration of a $\langle 110 \rangle$ self-interstitial atom defect.

\section{Methodology}\label{s:method}
\subsection{Simulation setup}

We performed DFT calculations using the OpenMX package \cite{openmx}, which implements pseudopotentials and pseudo-atomic orbitals. The bulk (pristine) simulation cell of Fe is constructed using $4 \times 4 \times 4$ unit cells  containing 128 atoms. We use the PBE generalised gradient approximation exchange correlation functional \cite{PBE,PBE2}, which together with the difference Hartree potential \cite{Ozaki_prb_2003} are evaluated on a real space grid. Numerical integration of these non-local terms are performed upon the discrete real-space grid partitioned by a cut-off energy of 600~Ry.

The basis set is created via a linear combination of optimised pseudo-atomic orbitals (LCPAO)  \cite{Ozaki_prb_2003,Ozaki_prb_2004,Ozaki_prb_2005}, employing three $s$, three $p$ and three $d$ orbitals centred on each atomic site, which all share a cut-off radius of 6 Bohr radii. Two-centre integrals in the Kohn-Sham Hamiltonian evaluated in momentum space use $3\times 3 \times 3$ k-points constructed by the Monkhorst-Pack method \cite{MP}. We use the Fe pseudopotentials of the form of Morrison, Bylander and Kleinman (MBK) \cite{Blochl_prb_1990,MBK} available within the OpenMX library, which include a non-linear partial core correction. The separable form of the MBK pseudopotentials are particularly suited for efficient LCPAO calculations. Ionic positions are relaxed until the maximum ionic force is smaller than $2\times 10^{-4}$ Ry/Bohr radius. 

We performed benchmark tests against literature data. We calculated the lattice constants, elastic constants and point defects formation energy and compared with data calculated by VASP \cite{KressePRB1993,KressePRB1994,KresseCMS1996,KressePRB1996} using projector augmented wave (PAW) potential \cite{Ma_PRM_2019} and ultrasoft pseudopotential (USPP) potential \cite{olsson_prb_2007}. We checked the convergence of our data against the k-points density, electronic temperature, and real-space cut-off energy. Bulk properties of BCC and FCC phases are presented in Section \ref{ss:bulk}. Defect formation energies are presented in Section \ref{ss:sia}. They all show good compatibility and confirm the validity of our results. 

\subsection{Exchange coupling parameter}

The LKAG equation \cite{LKAG_jmmm_1987} enables one to directly extract the scalar bilinear Heisenberg exchange integral $J_{ij}^{\text{eff}}$ from electronic structure calculations. Here we address the Green's function formalism of the LKAG equation \cite{Pajda_2001,Turek_2006}.
\begin{equation}\label{eqn:lkag1}
J_{ij}^{\text{eff}}(\mathbf{k}) = \frac{1}{\pi} \text{Im} \int^{\epsilon_F}_{-\infty} \text{Tr}\big(G^{\uparrow \uparrow}_{\mathbf{k},ij}\hat{V}^{\downarrow \uparrow}_{\mathbf{k},i}G^{\downarrow\downarrow}_{\mathbf{k},ji}\hat{V}^{\uparrow\downarrow}_{\mathbf{k},j}\big)d\epsilon.
\end{equation}
The single-particle Greens function at a given energy $\epsilon$ is defined by the resolvent of the Kohn-Sham orbitals $|\phi^{\sigma}_{\mathbf{k},i}\rangle$ in momentum space $\mathbf{k}$ over the filled states:
\begin{equation}\label{eqn:greens}
  G^{\sigma\sigma}_{\mathbf{k},ij}(\epsilon)=\displaystyle \sum _n \frac{|\phi^{\sigma}_{\mathbf{k},i}\rangle \langle \phi^{\sigma}_{\mathbf{k},j}|}{\epsilon - \epsilon^{\sigma}_{\mathbf{k},n}+i\eta}  
\end{equation}
where the spin index $\sigma$ for our collinear calculations refers to the majority $\uparrow$ and minority $\downarrow$ spins, $\epsilon^{\sigma}_{\mathbf{k},n}$ is the nth eigenenergy and $\eta$ is a positive infinitesimal smearing factor, implying the limit $\eta \rightarrow 0$. 

The philosophy leading to the derivation of the LKAG equation presents that a good and convenient way to determine the electronic structure of a system is to work within the Grand Canonical Potential (GCP). One is then able to relate variations in the GCP to changes in the integrated density of states. In turn, by use of Lloyd's formula \cite{Lloyd1972}, the integrated density of states may be expressed by a transition matrix which relates states of the perturbed system to the states of the unperturbed Hamiltonian. This may be represented as an open Born series constructed from successive expansions of retarded Green's functions of the unperturbed states (Eqn.~\ref{eqn:greens}) and an on-site scattering potential ($\hat{V}^{\uparrow\downarrow}_i$, Eqn.~\ref{eqn:onsite}). As a result, changes in the GCP owing to a small spin rotation can ultimately be determined just by knowing the relevant on-site potential. This is taken to be the potential difference induced by the rotation of the magnetic moment.

For collinear spins, the off-diagonal components of a local Hamiltonian $H^{\uparrow\downarrow}_i$ and $H^{\downarrow\uparrow}_i$ representing a given atomic site $i$ are zero. It follows that the on-site exchange splitting potential $\hat{V}^{\uparrow\downarrow}_{\mathbf{k},i}$ at atomic site $i$ due to an infinitesimal spin rotation can then be approximated using the difference in the local Hamiltonian between the up and down spin channels \cite{Pajda_2001,Han_prb_2004}. 
\begin{equation}\label{eqn:onsite}
    \hat{V}^{\uparrow\downarrow}_{\mathbf{k},i}=\frac{1}{2}\bigg(\hat{H}_{\mathbf{k},i}^{\uparrow\uparrow}-\hat{H}_{\mathbf{k},i}^{\downarrow\downarrow}\bigg).
\end{equation}

The local Hamiltonian is the partial matrix of the full Kohn-Sham Hamiltonian representing site $i$. Since our LCPAO calculations are using three $s$, three $p$ and three $d$ orbitals per Fe atom, the local Hamiltonian matrices have $27 \times 27$ matrix elements per spin state accessed via orbital indices.

Finally, LKAG is able to relate the changes to the Grand Canonical Potential to the bilinear exchange parameters by means of Eqn~\ref{eqn:lkag1}. We do not provide the derivations here and defer to the original publications \cite{LKAG_jmmm_1987, Bruno_prl_2003}. We present now the practical implementation of the LKAG as used in the following work.

For non-orthogonal LCPAO basis used in OpenMX, the MFT within the rigid spin approximation with non-collinear magnetic perturbations can be re-expressed in a practical manner as shown in Ref \cite{Han_prb_2004} by Han \emph{et al}. More recently, the same expression has been re-derived using local projection operators \cite{Steenbock_jctc_2015}. In the orbital representation

\begin{align}\label{eqn:jomx}
J_{ij}^{\text{eff}}(r_{ij}) &= \frac{1}{4} \int d\mathbf{k}\displaystyle \sum_{\alpha,\beta}
\bigg( \frac{f_{\alpha}^{\uparrow}-f_{\beta}^{\downarrow}}{\epsilon^{\downarrow}_{\beta}-\epsilon^{\uparrow}_{\alpha}+i\eta}\bigg) \nonumber \\
& \times 
\displaystyle \sum _{a,b}^{N_i} C^{\uparrow}_{\mathbf{k},\alpha a}\hat{V}_{ab}^{\downarrow\uparrow}C_{\mathbf{k},\beta b}^{\downarrow}
\displaystyle \sum _{a',b'}^{N_j} C_{\mathbf{k},\alpha a'}^{\uparrow}\hat{V}_{a'b'}^{\uparrow\downarrow}C_{\mathbf{k},\beta b'}^{\downarrow}
\end{align}
where the indices $a$ and $b$ run over the pseudoatomic orbitals centred on site $i$, and $a',b'$ span over site $j$. $\alpha, \beta$ are indices spanning all the orbitals in the system.  $f_{\alpha}^{\uparrow}$ and $f_{\beta}^{\downarrow}$ are the Fermi distributions
\begin{equation}
 f^{\sigma}_{\alpha}=\frac{1}{1+\exp ((\epsilon^{\sigma}_{\alpha}-\mu)/k_BT)} 
\end{equation}
with electron smearing temperature $T_e$ and chemical potential $\mu$. $\hat{V}_{ab}$ is the matrix element for the on-site potential at site $i$ between the orbitals centred at that site indexed $a$ and $b$. $C^{\sigma}_{\alpha a}$ are the molecular orbital coefficients of the self consistently solved generalised Kohn-Sham equations
\begin{equation}
\mathbf{HC}_{\alpha}=\epsilon_{\alpha}\mathbf{SC}_{\alpha}    
\end{equation}
where $\mathbf{C}_{\alpha}=(C_{\alpha 1},C_{\alpha 2},...,C_{\alpha N_{i}})^{\text{T}}$. This vector is constructed using a L\"{o}wdin transformation with the unitary vectors $\mathbf{U}$ that diagonalise the overlap of the Kohn-Sham orbitals $\mathbf{S}$. The corresponding eigenvalues $\mathbf{e}$ are necessarily positive definite. The transformation is then expressed as:
\begin{equation}\label{eqn:moc}
    \mathbf{C}_{\alpha}  = \frac{1}{\sqrt{e_{\alpha}}}\mathbf{U}_{\alpha}^{\dagger}
    \mathbf{H}\mathbf{U}_{\alpha}\frac{1}{\sqrt{e_{\alpha}}}.
\end{equation}
Since Eqn~\ref{eqn:moc} is a matrix equation, the identical positive definite terms involving the inverse square root of the overlap eigenenergies are non-commutative.

The exchange coupling parameter $J_{ij}^{\text{eff}}$ can then be calculated within the framework of DFT. OpenMX \cite{openmx} provides a utility that calculates $J_{ij}^{\text{eff}}$. However, it was primarily developed for calculation of molecules. Instead of using it, we developed our own code that has been optimised for bulk materials. We note that during the preparation of this manuscript a new release of OpenMX became available containing improvements to the exchange coupling code \cite{Terasawa_jphyssocjap_2019}.

In order to treat the variable magnitude of atomic spin, which we discuss below, we define another Heisenberg Hamiltonian $\mathcal{H}$
\begin{equation}\label{eqn:hhss}
    \mathcal{H} =- \displaystyle \sum _{i \neq j} J_{ij} \mathbf{S}_i\cdot \mathbf{S}_j
\end{equation}
where  $\mathbf{S}_i$ is the atomic spin vector at site $i$ and $J_{ij}$ is the exchange coupling parameter. In the Heisenberg Hamiltonian defined in Eq. \ref{eqn:hhe}, the magnitude of the atomic spins are subsumed into $J_{ij}^{\text{eff}}$. Comparing Eq. \ref{eqn:hhe} and \ref{eqn:hhss}, we see that the two exchange parameters are related by
\begin{equation}\label{eqn:jeff2j}
     J_{ij}  =  \frac{g^2\mu_B^2}{M_i M_j}J^{\text{eff}}_{ij}.
\end{equation}
The magnetic moments $\mathbf{M}_i$ can be determined by a Mulliken population analysis of the electronic density and overlap matrices:
\begin{equation}
    M_{\sigma,i\alpha} = \displaystyle \sum_n \displaystyle \sum_{j\beta} \rho ^{(\mathbf{R}_n)}_{\sigma,i\alpha j\beta} S_{i\alpha j\beta}^{(\mathbf{R}_n)}
\end{equation}
%
where $\rho^{(\mathbf{R_n})}_{\sigma,i\alpha,j\beta}$ is the density matrix pertaining to the periodic image of the simulation cell whose origin is positioned at $\mathbf{R_n}$. $S^{(\mathbf{R_n})}_{i\alpha,j\beta}$ is the overlap matrix. Indices $i$ and $j$ refer to the atomic sites, $\alpha$ and $\beta$ are the orbital indices, $\sigma$ denotes spin and $n$ spans the periodic images of the simulation cell within a given cutoff radius.

\section{Results}

\subsection{Bulk Iron}\label{ss:bulk}

We compare our bulk Fe data with other DFT calculations. The data in Table~\ref{tab:bulk}, produced using OpenMX calculations, show excellent agreement with other similar studies. The ground state of iron is expectantly found as the BCC ferromagnetic ($\alpha$) phase with atomic magnetic moments of 2.22$\mu_B$, in agreement with experiment \cite{Kittel2004}.

\begin{table}
\caption{\label{tab:bulk}Ground-state properties of BCC Fe as calculated using the OpenMX \cite{openmx}, comparing with plane wave DFT calculations using VASP \cite{Ma_PRM_2019} and experiment.}
\begin{ruledtabular}
\begin{tabular}{ c c c c }
Property                        & OpenMX       & VASP  & Exp   \\
                                & (Present) & \cite{Ma_PRM_2019}  &       \\ \hline
$a_0$ (\AA)                     & 2.842  & 2.831 & 2.87\cite{Kittel2004}\\
$\langle M \rangle$ $(\mu_B)$   & 2.25   & 2.21 & 2.22 \cite{Pepperhoff2010}\\
$\Omega_0$ (\AA$^3$)            & 11.49  & 11.34    & 11.82\cite{Kittel2004} \\
$C_{11}$ (GPa)                  & 242.12 & 289.34  & 243.1\cite{rayne_physrev_1961} \\
$C_{22}$ (GPa)                  & 138.74 & 152.34 & 138.1\cite{rayne_physrev_1961} \\
$C_{44}$ (GPa)                  & 87.72  & 107.43  &  121.9\cite{rayne_physrev_1961} \\
\end{tabular}
\end{ruledtabular}
\end{table}

The equilibrium lattice parameter $a_0^{\text{DFT}} = 2.842${\AA} is slightly underestimated relative to experimental value $a_0^{\text{exp}} = 2.8665${\AA}. This is not unexpected as overbinding effects are relatively common in the context of DFT calculations. 

The lowest energy magnetic configuration in the FCC phase is double-layer antiferromagnetic (AFM2), which is 0.1eV higher in energy than the FM BCC $\alpha$ phase. They are consistent with Ref. \onlinecite{ma_prb_2017}. We notice a small discrepancy in the stability of the FCC magnetically ordered phases (Table~\ref{tab:FCC}). Our calculations find the next stable configurations to be the high-spin ferromagnetic (HS) and single layer antiferromagnetic (AFM), which are nearly degenerate at 0.12 and 0.13 eV, respectively. This differs from Ref. \cite{ma_prb_2017} where the stability was explored using the PAW method and where the AF1 phase was found to be the next stable phase, with the HS and ferromagnetic low spin (LS) configurations being of comparable stability. This difference likely arises from differences between the pseudopotentials used in the two approaches.

\begin{figure}
\includegraphics[width=8.5cm]{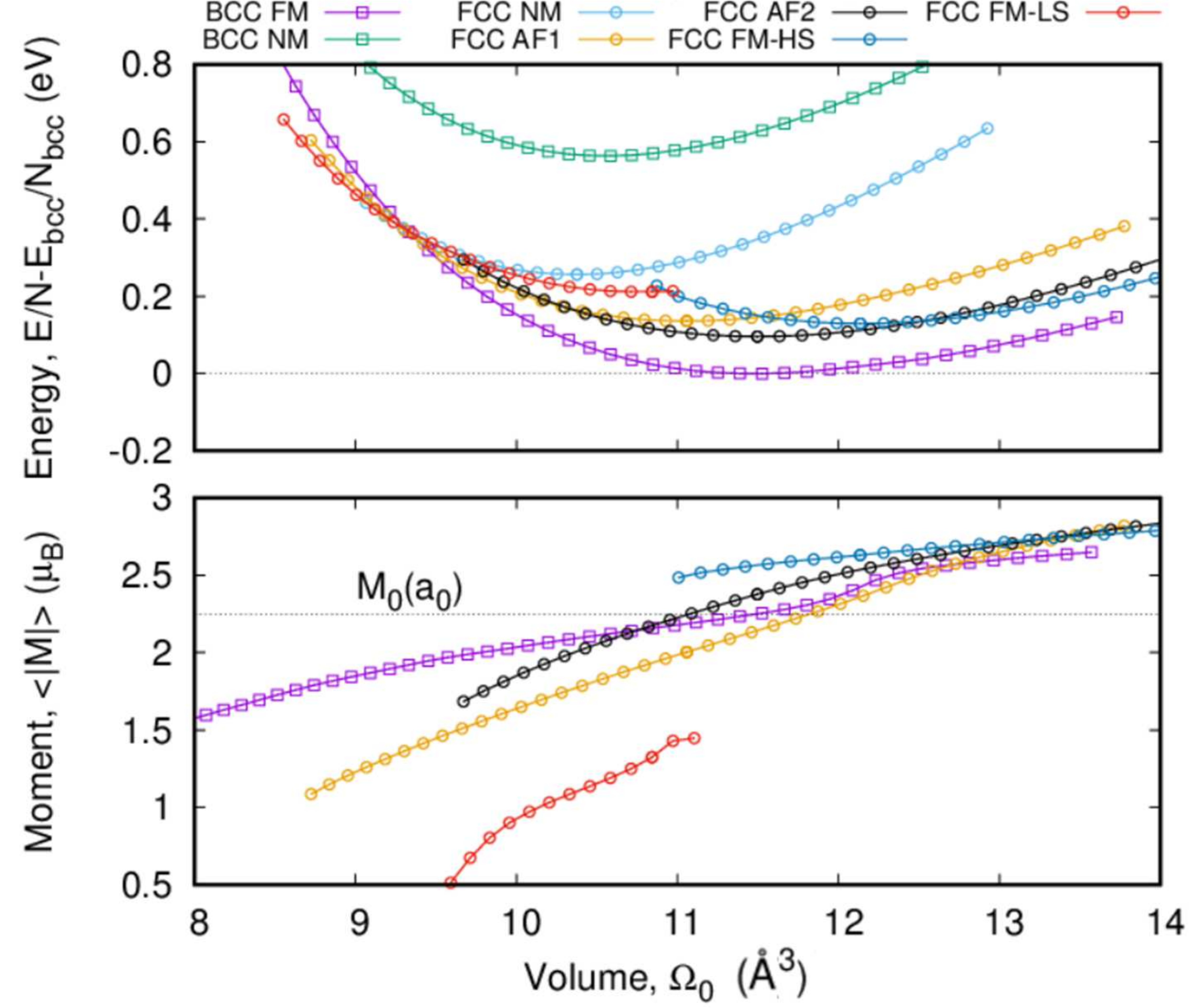}
\caption{\label{fig:bulk}(a) Energy and (b) magnitude of magnetic moment of different magnetic configurations of BCC and FCC Fe calculated using OpenMX. Energies are normalised per atom and are shown relative to the global 0K ground state (BCC FM). Atomic volume $\Omega_0$ is computed as the total volume of the simulation cell divided by the number of atoms in it, $|\vec{L_x}\times \vec{L_y} \cdot \vec{L_x}| / N$. The following configurations are shown: BCC ferromagnetic (FM), BCC non-magnetic (NM), FCC non-magnetic (NM), FCC high-spin (HS), FCC low-spin (LS), FCC antiferromagnetic (AF1) and FCC double layer antiferromagnetic (AF2).}
\end{figure}

In Fig. \ref{fig:bulk} we plot the magnitude of the magnetic moments as a function of volume for different magnetically ordered phases. We find quantitative agreement with previous calculations showing that the magnitude of the moments decreases under compression due to the increasing exchange energy to satisfy the Pauli exclusion principle. We also observe an inflection point in the $\alpha$ phase when, under tension, the lattice parameter of 1.014$a/a_0^{\text{DFT}}$ (where $\Omega_0\approx 12$ \AA$^3$ in Figure~\ref{fig:bulk}) is reached. This kinking is known to occur due to large changes in the $t_{{\text {2g}}}$ density of states at the Fermi level $\epsilon_F$ relative to smaller changes in the density of states associated with  the $e_{\text{g}}$ orbitals \cite{Wang_prb_2010}. 

\begin{table*}
\caption{\label{tab:FCC} Comparison of the ground state FCC Fe magnetic structures calculated with OpenMX with reference data. Atomic volumes $\Omega_0$ are given in {\AA}$^3$. Values in the last column represent the difference between the energy per atom computed for a given structure and the energy per atom in the ferromagnetic BCC phase. Values as functions of volume are plotted in Figure~\ref{fig:bulk}.}
\begin{ruledtabular}
\begin{tabular}{ c c c c c c}
Configuration   & $\Omega_0$ (\AA$^3$)  & Reference             & $\langle |M| \rangle$ ($\mu_B)$  & Reference         & Energy diff.  \\ 
                & (Present)             & (\AA$^3$)             & (Present)                      & ($\mu_B)$         & (eV)     \\ \hline
AFM1            & 11.05                 & 10.76, 11.37          & 2.00                           & 1.574             & 0.13     \\
AFM2            & 11.50                 & 11.20                 & 2.376                          & 2.062             & 0.096    \\
FM-HS           & 12.14                 & 11.97,12.12           & 2.631                          & 2.572             & 0.12     \\
FM-LS           & 10.84                 & 10.52                 & 1.324                          & 1.033             & 0.21     \\
NM              & 10.38                 & 10.22                 & 0.000                          & 0.000             & 0.25     \\
\end{tabular}
\end{ruledtabular}
\end{table*}

\begin{table}
\caption{\label{tab:jij} Values of effective exchange coupling parameters $J_{ij}^{\text{eff}}$ evaluated using the Magnetic Force Theorem. Values of $J^{\text{eff}}_{ij}$ were computed assuming the experimental lattice parameter $a_0^{\text{exp}}=2.8665${\AA} or the DFT equilibrium lattice parameter $a_0^{\text{DFT}}=2.842${\AA}  (in parenthesis). The Curie temperature $T_C$ can be estimated in the mean field approximation using Eq. \ref{eqn:Tc_mfa}. We included contributions from the four nearest neighbour shells, where in BCC case $T_C \approx 2(8 J^{(1)}+6 J^{(2)} + 12 J^{(3)} +24 J^{(4)})/3k_B$.}
\begin{ruledtabular}
\begin{tabular}{ c c c c c }
$J_{ij}^{\text{eff}}$   & LCPAO(GGA)        & LMTO       & LMTO       & LMTO \\ 
(mRy)                   & (Present)         & (GGA)\cite{Wang_prb_2010}      & (LSDA)\cite{frota_prb_2000}    & (LSDA)\cite{Katnelson_prb_2000}     \\ \hline
$J^{(1)}$         & 1.204 (1.14)      & 1.218     & 1.24      & 1.212 \\  
$J^{(2)}$        & 0.953 (0.72)      & 1.08      & 0.646     & 0.593 \\
$J^{(3)}$         & -0.035 (-0.004)   & -0.042    & 0.007     & 0.018 \\
$J^{(4)}$        & -0.085 (-0.087)    & -0.185    & -0.108    & -0.07 \\
\hline \hline
$T_{C}$ (K)             & 1362 (1193)          & 1186      & 1170      & 1240 \\  
\end{tabular}
\end{ruledtabular}
\end{table}

\begin{figure}
\includegraphics[width=8.5cm]{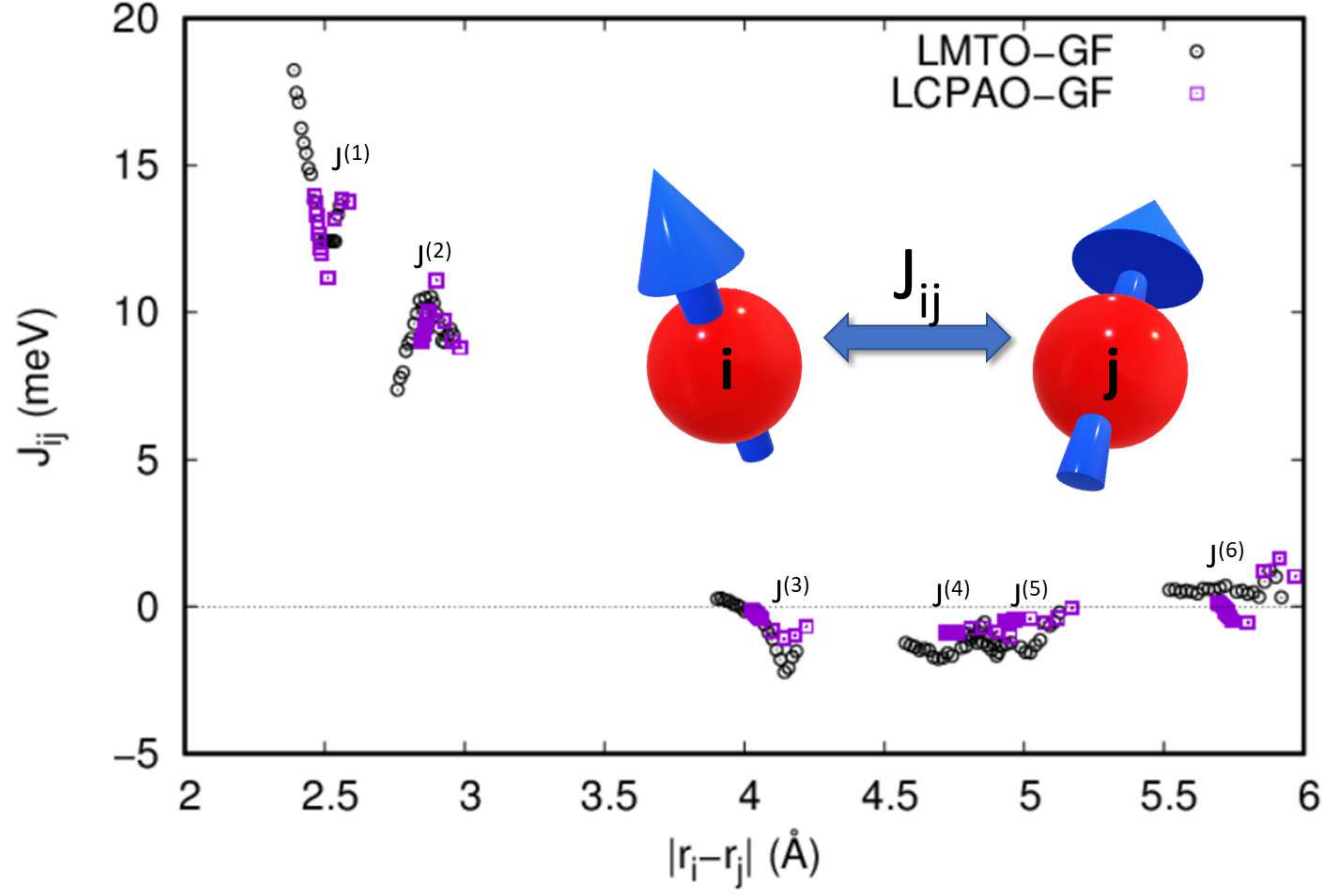}
\caption{\label{fig:jij_bulk} The exchange coupling parameter $J_{ij}$ as a function of interatomic distance. Black circles are data from Ref. \cite{ma_prb_2017}, calculated using the LMTO-GF method \cite{LKAG_jmmm_1987, Schilfgaarde_jap_1999}. Purple squares are the current results calculated using LCPAO and the LKAG method \cite{LKAG_jmmm_1987} (Eq. \ref{eqn:jomx}), where the volume of a simulation box containing 128 atoms varies such that the linear dimension change in the range of $\pm3\%$.
}
\end{figure}

Values of the exchange coupling parameter $J_{ij}^{\text{eff}}$ of BCC ferromagnetic Fe calculated using the LKAG equation \cite{LKAG_jmmm_1987} (Eq. \ref{eqn:jomx}) are given in Table~\ref{tab:jij}. The values were computed assuming the experimentally observed lattice parameter, or the lattice parameter corresponding to the DFT energy minimum (values given in parenthesis). The table gives the values of exchange parameters computed for the first four nearest neighbour shells, the corresponding values are denoted by  $J^{(1)}$, $J^{(2)}$, $J^{(3)}$ and $J^{(4)}$. 

Two recent studies performed using the LMTO \cite{Korotin_prb_2015} and LCPAO \cite{Yoon_PRB_2018} tested the dependence of the computed values of  exchange parameters on the choice of the basis set. Depending on the choice of basis functions, the calculated values of exchange parameters  $J_{ij}^{\text{eff}}$ can vary by 3meV (0.2mRy). It has also been noted that DMFT corrections affect the magnitude of orbitally resolved $J_{ij}^{\text{eff}}(r_{ij})$, but the sign and relative strength remains unaltered \cite{Kvashnin_prl_2016}. It suggests that whilst we could opt for a more sophisticated method, our results summarised in Table \ref{tab:jij} are informative and show good compatibility with the published data \cite{Wang_prb_2010,frota_prb_2000,Katnelson_prb_2000}. 

We explored the variation of the effective exchange coupling parameter $J_{ij}^{\text{eff}}(r_{ij})$ treated as a function of interatomic distance $r_{ij}$ by varying the volume of the simulation cell. The linear dimension of the cell varied in the range of $\pm3$\%. Fig. \ref{fig:jij_bulk} shows the calculated exchange coupling parameter $J_{ij}$ defined according to Eq. \ref{eqn:jeff2j}. Again, the data agree with the results from Ref. \cite{ma_prb_2017}, where the calculations were performed using the LMTO Green’s function technique, developed and implemented by van Schilfgaarde \textit{et al.} \cite{LKAG_jmmm_1987,Schilfgaarde_jap_1999}.

Using the values $J_{ij}^{\text{eff}}$ computed for several coordination shells, the Curie temperature $T_C$ can be estimated in the  mean field approximation \cite{LKAG_jmmm_1987} as
\begin{equation}\label{eqn:Tc_mfa}
k_B T_{C} \approx \frac{2}{3}J_0^{\text{eff}}
\end{equation}
where $J_0^{\text{eff}}= \sum_{j\neq 0}J_{0j}^{\text{eff}}$. The data given in Fig. \ref{fig:jij_bulk} show that $J^{(1)}$ and $J^{(2)}$ give the dominant contribution to $J_0^{\text{eff}}$. Still, we evaluate $J_0^{\text{eff}}$ using the effective exchange parameters for the coordination shells extending to the \nth{4} nearest neighbour. The estimated values of $T_C$ are given in Table~\ref{tab:jij} together with other values, taken from literature and also calculated in the mean field approximation.

\subsection{Point defects in Iron}\label{ss:sia}

We now investigate magnetic interactions in iron containing point defects, and compare the results to the bulk case. First, we benchmark the calculated formation energy of a self-interstitial atom (SIA) defect and a vacancy against literature data \cite{domain_prb_2001,fu_prl_2004,Willaime2005,olsson_prb_2007}. Then, we study how the exchange coupling parameters vary in the vicinity of a defect, especially near the core of a defect configuration.

The formation energy $E^{F}_{\text{def}}$ of a defect formed in a given structural and magnetic phase can be written as
\begin{equation}
    E^F_{\text{def}} = E_{\text{def}}(N_{\text{def}})-\frac{N_{\text{def}}}{N_{\text{bulk}}}E_{\text{bulk}}(N_{\text{bulk}}) 
\end{equation}
where $E_{\text{def}}(N_{\text{def}})$ is the energy of a system including the defect and $E_{\text{bulk}}(N_{\text{bulk}})$ is the energy of the reference perfect system. The number of atoms in each system is $N_{\text{def}}$ and $N_{\text{bulk}}$, respectively. For the cell size and defect structures considered here we ignore the elastic correction to the formation energy of the defect \cite{Ma_PRM_2019}, as the magnitude of the elastic correction varies between 0.2 and 0.3 eV whereas the variation of DFT parameters leads to an absolute error in the formation energy of the order of 0.05-0.1 eV per SIA \cite{Becquart_jnm_2018}, which is a quantity of similar magnitude. 

The simulation cell for a defect calculation is chosen to be of the same shape and volume as in the perfect lattice case. SIA configurations are created by inserting additional Fe atoms at different positions in the lattice, and all the ionic positions are then relaxed until all the forces acting on ions are lower than $2\times 10^{-4}$ Ry/Bohr radius. We considered the $\langle 100 \rangle$, $\langle 110 \rangle$ and $\langle 111 \rangle$ dumbbell defect configurations, a $\langle 111 \rangle$ crowdion, a tetrahedral site interstitial and an octahedral site interstitial. A vacancy configuration is created by removing an atom, followed by the relaxation of ionic positions.

\begin{table*}
\caption{\label{tab:efsia} Calculated defect formation energies $E^{F}_{\text{def}}$.  Calculations were performed using a LCPAO basis set. Our results are compared with Refs. \cite{domain_prb_2001,olsson_prb_2007,fu_prl_2004,Ma_PRM_2019} where calculations were performed using LCPAO, plane-wave with PAW, or plane-waves with USPP. Values in parenthesis show the energy difference of an SIA configuration with respect to the formation energy of a $\langle 110 \rangle$ dumbbell. Due to the short range of the LCPAO, we present the calculated formation energy of a vacancy with (daggered) and without the additional basis functions added to the vacancy site. All the calculations were performed using 128$\pm 1$ atom cells.}
\begin{ruledtabular}
\begin{tabular}{ c c c c c c }

        & PAO &  PAW & PAO & USPP & PAW \\
Defect  & (Present) & [\onlinecite{olsson_prb_2007}] & [\onlinecite{fu_prl_2004}] & [\onlinecite{olsson_prb_2007}] & [\onlinecite{Ma_PRM_2019}] \\ \hline
\hline
$\langle 100 \rangle_{\text{D}}$             & 5.58 (1.10)  & 5.13 (1.11) & 4.64 (1.00) & 5.04 (1.10) & 5.59 (1.17) \\
$\langle 110 \rangle_{\text{D}}$             & 4.49         & 4.02        & 3.64        & 3.94        & 4.42        \\
$\langle 111 \rangle_{\text{D}}$  & 5.26 (0.80)  &             & 4.34 (0.7)  & 4.66 (0.72) & 5.21 (0.79) \\
$\langle 111 \rangle_{\text{C}}$  & 5.27 (0.79)  & 4.72 (0.7)  &             &             & 5.21 (0.79) \\
Tetrahedral                        & 4.98 (0.49)  & 4.44 (0.42) & 4.26 (0.62) & 4.46 (0.52) & 4.88 (0.46) \\
Octahedral                        & 5.74 (1.25)  & 5.29 (1.27) & 4.94 (1.30) & 5.25 (1.31) & 5.68 (1.26) \\
Vacancy                           & 2.26 / 2.18$^{\dagger}$ & 2.15        & 2.07        & 2.02        & 2.19        \\
\end{tabular}
\end{ruledtabular}
\end{table*}

The formation energies of point defects in BCC Fe are summarised in Table \ref{tab:efsia}, and compared with results given in Refs. \cite{olsson_prb_2007,fu_prl_2004,Ma_PRM_2019}. The most stable SIA configuration is the $\langle 110 \rangle$ dumbbell, which agrees with earlier results \cite{domain_prb_2001,fu_prl_2004,nguyen-manh_prb_2006,derlet_prb_2007}. The relative stability also follows the same order, such that the formation energies are ordered as $\langle 110 \rangle_D$ $<$ tetrahedral  $<$ $\langle 111 \rangle_C$ $<$ $\langle 111 \rangle_D$ $<$ $\langle 100 \rangle_D $ $<$ octahedral, where the corresponding configurations are 0.49, 0.79, 0.81, 1.10 and 1.25 eV higher in energy that the $\langle 110 \rangle$ dumbbell, respectively. Our results agree well with the literature data derived using different basis sets and pseudopotentials.

The magnetic moments of SIA configurations are also consistent with those reported in literature \cite{domain_prb_2001,olsson_prb_2007}. In general, the magnetic moments in the core of an SIA configuration are significantly suppressed. Moments at the tensile \nth{1} n.n. sites are enhanced whilst those at sites characterised by a compressive strain are slightly decreased relative to the bulk value. A more complex relation between the local structure and local magnetic moment is found in C15 defect clusters \cite{Marinica2012}.

In the case of a $\langle 110 \rangle$ dumbbell, the magnetic moments of the two Fe atoms at the core of the defect are antiparallel with respect to the  surrounding atoms, and the magnitude of both moments is -0.30$\mu_B$. This is slightly larger than what is found in calculations performed using the PAW method, which predicts the value of -0.1$\mu_B$ \cite{Ma_PRM_2019} and USPP, which gives -0.2$\mu_B$ \cite{olsson_prb_2007}. 

In the case of a $\langle 111 \rangle$ dumbbell, the two core atoms are in the ferromagnetic state having moments of +0.19$\mu_B$. In Ref. \cite{olsson_prb_2007}, the two core atoms can be ferromagnetic (0.3$\mu_B$) or antiferromagnetic (-0.5$\mu_B$), if USPP or PAW methods is used, respectively. These results are in good agreement with literature data. We now move on to the calculations of the exchange coupling parameters. 

\begin{figure}
\includegraphics[width=8.5cm]{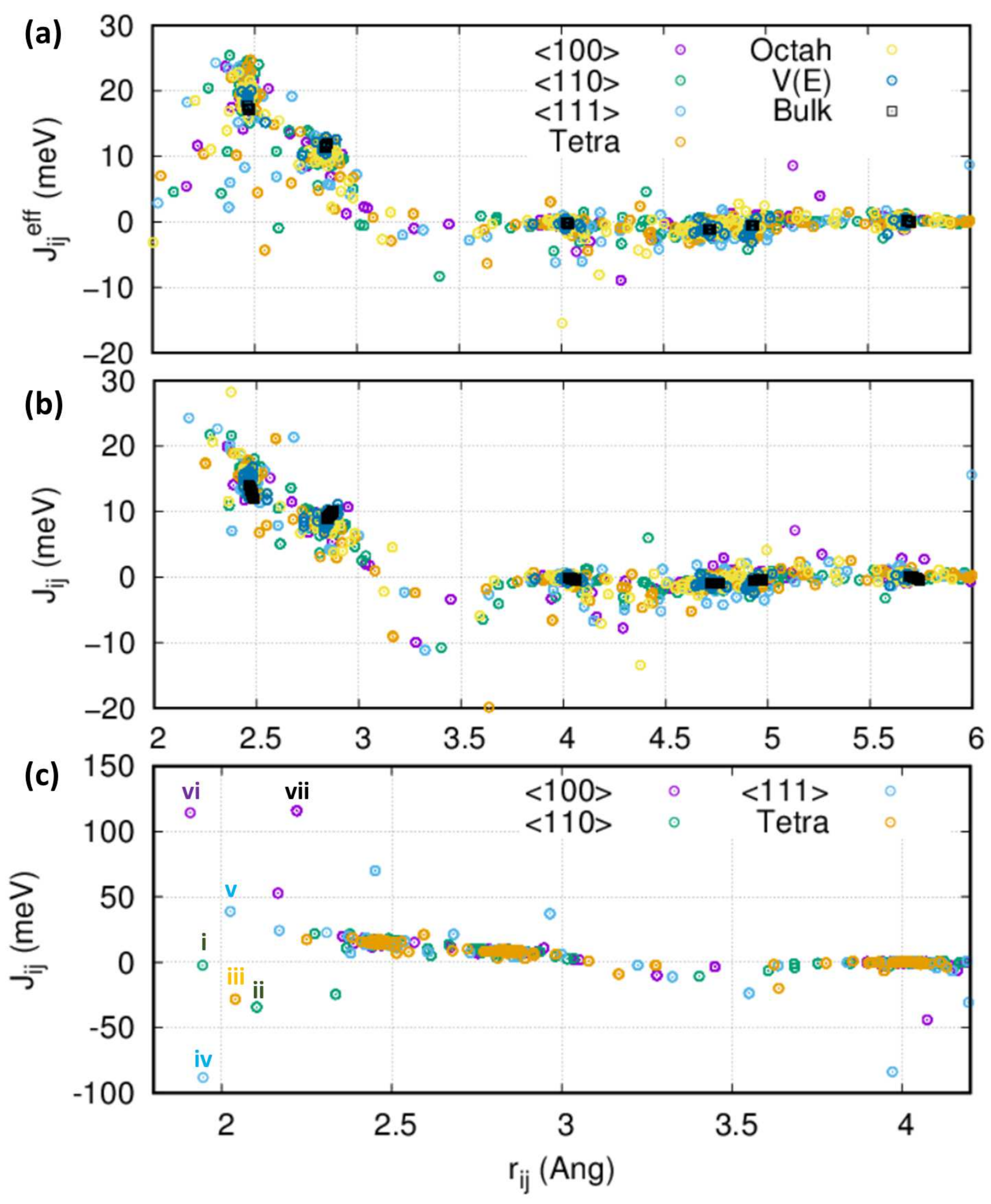}
\caption{\label{fig:jij_sia} (a) $J^{\text{eff}}_{ij}$ (Eq. \ref{eqn:jomx}) and (b) $J_{ij}$ (Eq. \ref{eqn:jeff2j}) as a function of interatomic distance $r_{ij} = |\mathbf{r}_i-\mathbf{r}_j|$ for SIA defects, including $\langle 100 \rangle$, $\langle 110 \rangle$, $\langle 111 \rangle$ dumbbells, a tetrahedral site interstitial and an octahedral site interstitial, and a vacancy configuration. Bulk values are also shown for comparison. (c) $J_{ij}$ shown over a greater ordinate range to include the outliers.}
\end{figure}


In Fig. \ref{fig:jij_sia}, we present both the effective exchange coupling parameter $J^{\text{eff}}_{ij}$ (Eq. \ref{eqn:jomx}) and the scaled exchange coupling parameter $J_{ij}$ (Eq. \ref{eqn:jeff2j}) computed for various SIA and vacancy configurations. $J^{\text{eff}}_{ij}$ at around the perfect lattice \nth{1} n.n. distance has the magnitude in the range of 15-25meV, whereas its value in a perfect lattice is 19meV. Both $J^{\text{eff}}_{ij}$ and $J_{ij}$  tail off quickly by the \nth{3} n.n., where $r_{ij}\approx 4$\AA. The exchange coupling parameter depends on the overlap between the localised basis functions, and so decays rapidly. Despite overlapping with orbitals of the core atoms, the values of $J^{\text{eff}}_{ij}$ do not change much for bulk-like atoms surrounding the defect core. One can expect the bulk exchange coupling parameters to be a good approximation to them.

\begin{table*}
\caption{Exchange coupling parameters between a core atom ($\alpha$) and its \nth{1} nearest neighbours ($\beta$). Multiplicity is defined per $\alpha$ atom. The key identifies specific data point plotted in Fig. \ref{fig:jij_sia}c. We indicate whether the exchange coupling parameter contributes to increase ($\uparrow$) or decrease ($\downarrow$) the energy of the system. }
\begin{ruledtabular}
\begin{tabular}{ c c c c c c c c }
\label{tab:je}
SIA                   & Key  & Pair                & $M$ ($\mu_B$)         & Order & Multiplicity & $J^{eff}$ (meV) & $J$ (meV)  \\ \hline \hline
$\langle 110 \rangle_D$ & i    & $J_{\alpha \alpha}$ & $M_{\alpha}=-0.304$   & FM    & 1 & -0.05 & -2 ($\uparrow$) \\
                      & ii   & $J_{\alpha \beta}$  & $M_{\beta} =+1.75$  & AFM   & 2 & 19.6  & -35 ($\downarrow$) \\ \hline
Tetrahedral           & iii  & $J_{\alpha \beta}$  & $M_{\alpha}=-0.87$  & AFM   & 4 & 6.8   & -28($\downarrow$) \\
                      &      &                     & $M_{\beta}=+1.13$   &      &   &       & \\ \hline
$\langle 111 \rangle_D$ & iv   & $J_{\alpha \alpha}$ & $M_{\alpha}=+0.198$ & FM    & 1 & -0.84 & -88 ($\uparrow$) \\
                      & v    & $J_{\alpha \beta}$  & $M_{\beta}=+1.49$   & FM    & 1  & 3.00  & +38 ($\downarrow$) \\ \hline
$\langle 100 \rangle_D$ & vi   & $J_{\alpha \alpha}$ & $M_{\alpha}=+0.188$ & FM    & 1 & 1.0   & +114 ($\downarrow$) \\
                      & vii  & $J_{\alpha \beta}$  & $M_{\beta}=+2.14$   & FM    & 4 & 11.6  & +116 ($\downarrow$) \\
\end{tabular}
\end{ruledtabular}
\end{table*}

Here we introduce the notation $\alpha, \beta$ and $\gamma$ as dummy indices representing the index of the core atoms, and their \nth{1} and \nth{2} n.n., respectively. For the \nth{1} and \nth{2} n.n. of the core atoms, where there is a greater degree of orbital overlapping,  $J^{\text{eff}}_{ij}$ behaves like in a glassy material with scattered values $-0.5 < J^{\text{eff}}_{\alpha \beta} < 27$meV. If we look at $J_{ij}$ (Eq. \ref{eqn:jeff2j}) instead, the magnitude of $J_{\alpha \alpha}$ and $J_{\alpha \beta}$ are 2-5 times greater than for bulk \nth{1} n.n. interaction, as shown in Fig. \ref{fig:jij_sia} c. 

In Table \ref{tab:je}, values of parameters $J_{\alpha \alpha}$ and $J_{\alpha \beta}$ are presented for the $\langle 100 \rangle$, $\langle 110 \rangle$, $\langle 111 \rangle$ dumbbells and a tetrahedral site interstitial. They are also shown in Fig.~\ref{fig:jij_sia}c via the key indexes. We can understand that the suppression of the magnetic moment of the core atoms is responsible for the large increase in $J_{ij}$ between the core and proximate neighbours, which occurs due to the fact that $J_{ij}\propto 1/M_iM_j$. 

Exchange coupling may increase or decrease the energy of a system. According to the definition of the Heisenberg Hamiltonian (Eq \ref{eqn:hhss}),  aligned spins ($\mathbf{M}_i\cdot \mathbf{M}_j > 0$) with $J_{ij} > 0$ will lower the energy. On the other hand, if  $J_{ij} < 0$, the antiparallel orientation of moments is favourable. In most cases, the magnetic energy of $\alpha-\alpha$ and $\alpha-\beta$ exchange interactions acts to lower the energy of the system. The exception is the $\langle 111 \rangle$ dumbbell. The magnetic interaction between the two core atoms contributes +0.1 eV per atom. Overall, the magnetic interactions lower the total energy of the $\langle 111 \rangle$ dumbbell (and the corresponding crowdion, not shown). However, the positive contribution from the repulsion between the core atoms causes the total energy to reduce by less than for other SIA configurations. This raises the relative energy of defect configurations, directly changing the order of their stability. It agrees with previous studies \cite{nguyen-manh_prb_2006,Ma_PRM_2019} suggesting that magnetism is the cause making the $\langle 110\rangle_D$ configuration more stable than the $\langle 111\rangle_D$ configuration. 

\subsection{Failure of the Heisenberg Hamiltonian}\label{ss:fail}

For a fixed atomic configuration $\mathcal{R}=\{\mathbf{r}_i\}$, we may calculate the energy change due to a specific spin ordering $\mathcal{S}=\{\mathbf{S}_i\}$. We may define the magnetic contribution to energy as the difference between the magnetic and non-magnetic states of configuration $\mathcal{R}$:
\begin{equation}\label{eqn:dft_cont}
    E_{\text{MC}}^{\text{DFT}}(\mathcal{R},\mathcal{S}) 
    = E^{\text{DFT}}_{\text{M}}(\mathcal{R},\mathcal{S})-E^{\text{DFT}}_{\text{NM}}(\mathcal{R})
\end{equation}
where $E^{\text{DFT}}_{\text{M}}$ and $E^{\text{DFT}}_{\text{NM}}$ are the cohesive energies calculated with and without spin polarisation from DFT, respectively. 

Using the LKAG equation\cite{LKAG_jmmm_1987} we aim to map the DFT magnetic energy contribution onto the Heisenberg functional form. We note the the LKAG equation is derived using relations from the second derivative of the energy with respect to the atomic spin

\begin{align}
    \frac{\partial ^2 E_{\text{MC}}^{\text{DFT}}}{\partial \mathbf{S}_i \partial \mathbf{S}_j}
    =\frac{\partial ^2 E_{\text{M}}^{\text{DFT}}}{\partial \mathbf{S}_i \partial \mathbf{S}_j}
    \approx -J_{ij}
\end{align}

If $E^{\text{DFT}}_{\text{MC}}(\mathcal{R},\mathcal{S})$ varies approximately the same as the Heisenberg Hamiltonian $E_{\text{MC}}^{\text{HH}}$, the system can be said to be a good \emph{Heisenberg magnet}. 

In Table \ref{tab:energy}, we list out the contributions of each term in Eq. \ref{eqn:dft_cont}. The non-magnetic calculations were performed using the relaxed atomic configuration from the corresponding spin-polarised calculation. The use of non-spin-polarised calculation changes the order of stability of SIAs. It decreases the energy of the $\langle 111\rangle_D$ relative to the tetrahedral site interstitial. This is consistent with the values of exchange coupling parameters between the core atoms and their neighbours of the $\langle 111\rangle_D$, which increases the energy (Table~\ref{tab:je}). 

On the other hand, the energy of the $\langle 100 \rangle$ dumbbell in the non-magnetic calculations increases relative to the $\langle 110 \rangle$ by 0.3eV. Despite the 0.1eV magnetic contributions between each of the $\alpha$ and $\beta$ ions, which lower the energy significantly, the $\langle 100 \rangle$ dumbbell remains energetically unfavourable. 

As a brief note, one may consider the relaxation of defect structures directly using non-spin-polarised DFT to determine the non-magnetic order of stability. However, we find that these structures have negative tetragonal shear modulii and are therefore mechanically unstable.
\begin{table*}
\caption{Calculated values of various energy terms computed for simulation cells containing relaxed SIA configurations (in Rydberg units). We record as cohesive energies, shifted relative to the energy of the ground state BCC structure ($E=E^{\text{calc}}-N_{atom}E^{\text{BCC}}_{\text{ref}}$). Energy differences with the $\langle 110 \rangle$ configuration are given in parenthesis in units eV. Subscripts correspond to the following nomenclature: M=Cohesive energy in a spin-polarised calculation, NM=Cohesive energy in a non-spin-polarised calculation, MC=Magnetic Contribution, HH=Heisenberg Hamiltonian and HL=Heisenberg-Landau Hamiltonian. Further, we may define the energy terms as: $E_{\text{MC}}^{\text{DFT}}=E^{\text{DFT}}_{\text{M}}-E^{\text{DFT}}_{\text{NM}}$, 
$E_{\text{M}}^{\text{HH}}=E^{\text{DFT}}_{\text{NM}}+E^{\text{HH}}_{\text{MC}}=E^{\text{DFT}}_{\text{NM}}+\mathcal{H}$, 
$E_{\text{M}}^{\text{HL}}=E^{\text{DFT}}_{\text{NM}}+\mathcal{H}_L$}
\begin{ruledtabular}
\begin{tabular}{ c  c c c c }
\label{tab:energy}
& $\langle 110 \rangle_D$ 
& Tetrahedral
& $\langle 111 \rangle_D$
& $\langle 100 \rangle_D$ \\ \hline\hline
$E^{\text{DFT}}_{\text{M}}$       & 0.3298  &  0.3657 ($\Delta_{110}$=0.49eV)  & 0.3879 ($\Delta_{110}$=0.79eV) & 0.4107 ($\Delta_{110}$=1.09eV)\\
$E^{\text{DFT}}_{\text{NM}}$      & 5.9319  &  5.9691 ($\Delta_{110}$=0.51eV)  & 5.9604 ($\Delta_{110}$=0.38eV) & 6.0359 ($\Delta_{110}$=1.40eV) \\
$E^{\text{DFT}}_{\text{MC}}$      &-5.6021  & -5.6033           &-5.5725          & -5.6252 \\ \hline
$E^{\text{HH}}_{\text{MC}}$       &-1.7206  & -1.7167 ($\Delta_{110}$=0.06eV)  &-1.7071 ($\Delta_{110}$=0.18eV) & -1.7423 ($\Delta_{110}$=-0.30eV) \\
$E^{\text{HH}}_{\text{M}} \quad$  & 4.2113  &  4.2524 ($\Delta_{110}$=0.56eV)  & 4.2534 ($\Delta_{110}$=0.57eV) & 4.2935 ($\Delta_{110}$=1.12eV)\\ \hline
$\sum A_i \mathbf{S}_i^2$         &-7.7630 & -7.7735          &-7.7318         &-7.7666\\
$\sum B_i \mathbf{S}_i^4$         & 5.6021  & 5.6034           & 5.5730          & 5.6257\\
$E_{\text{M}}^{\text{HL}}$        & 0.3298  & 0.3657  ($\Delta_{110}$=0.49 eV) & 0.3875 ($\Delta_{110}$=0.79 eV) & 0.4103 (1.09 eV)\\ \hline
$E^{\text{DFT}}_{\text{M}}$-$E_{\text{M}}^{\text{HL}}$ & 0.000 & 0.000 & 0.004 & 0.004 \\
\end{tabular}
\end{ruledtabular}
\end{table*}

Provided that $E^{\text{DFT}}_{\text{MC}} \approx E_{\text{MC}}^{\text{HH}}$, one may approximate the total magnetic energy using the Heisenberg Hamiltonian as
\begin{equation}
E^{\text{DFT}}_{\text{M}}(\mathcal{R},\mathcal{S}) \approx E_{\text{MC}}^{\text{HH}}(\mathcal{R},\mathcal{S})  + E^{\text{DFT}}_{\text{NM}}(\mathcal{R}),
\end{equation}
as one may wish to achieve in a multiscale model.

From Table \ref{tab:energy}, we observe that $E_{\text{MC}}^{\text{HH}}$ and $E_{\text{MC}}^{\text{DFT}}$ differ by about 4Ry. In terms of the relative energy difference with respect to the $\langle 110\rangle_D$, the most deviated case is the $\langle 111\rangle_D$. From DFT we expect it to be $+0.79$~eV higher in energy than the $\langle 110 \rangle_D$ configuration but we find that the energy difference is $+0.57$~eV from the Heisenberg Hamiltonian.

\begin{figure}
\includegraphics[width=8.5cm]{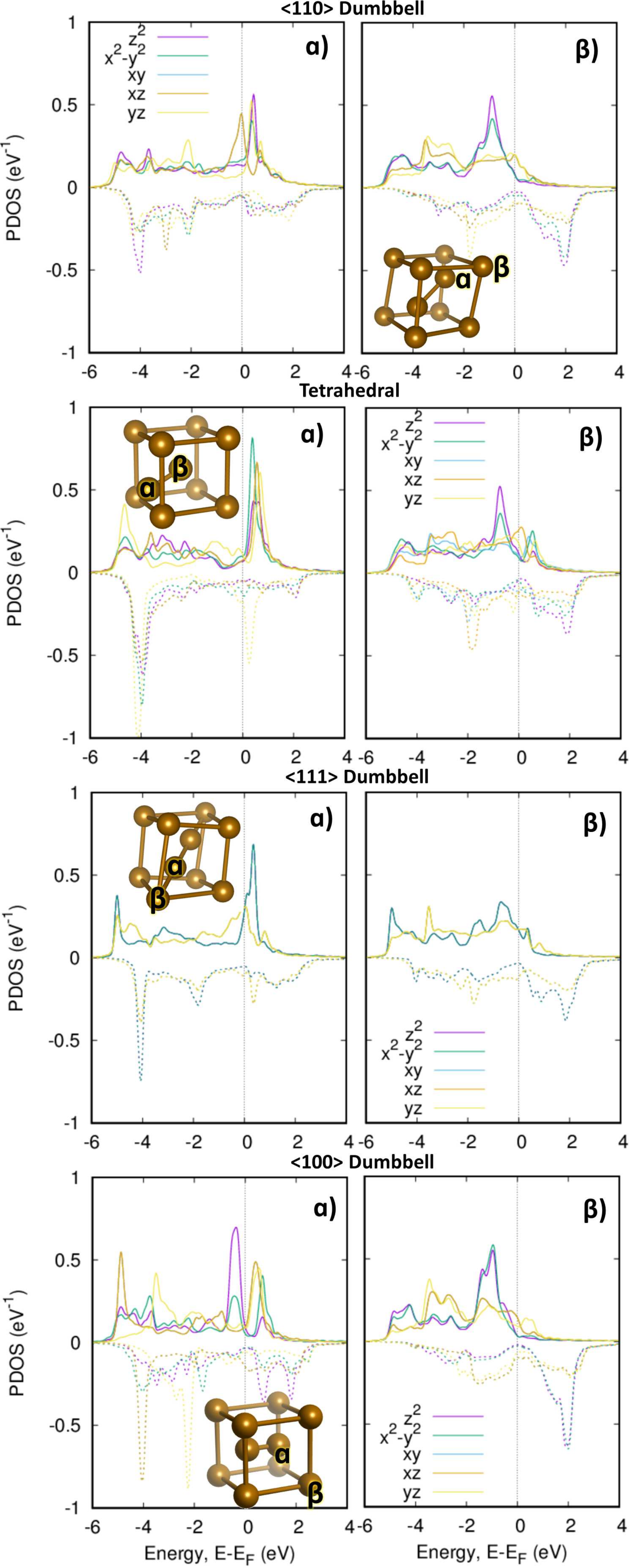}
\caption{\label{fig:pdos} Orbitally projected local density of states (PDOS) calculations for the core atoms ($\alpha$) and their nearest neighbours ($\beta$) of various SIA configurations.}
\end{figure}

\begin{figure}
\includegraphics[width=8.5cm]{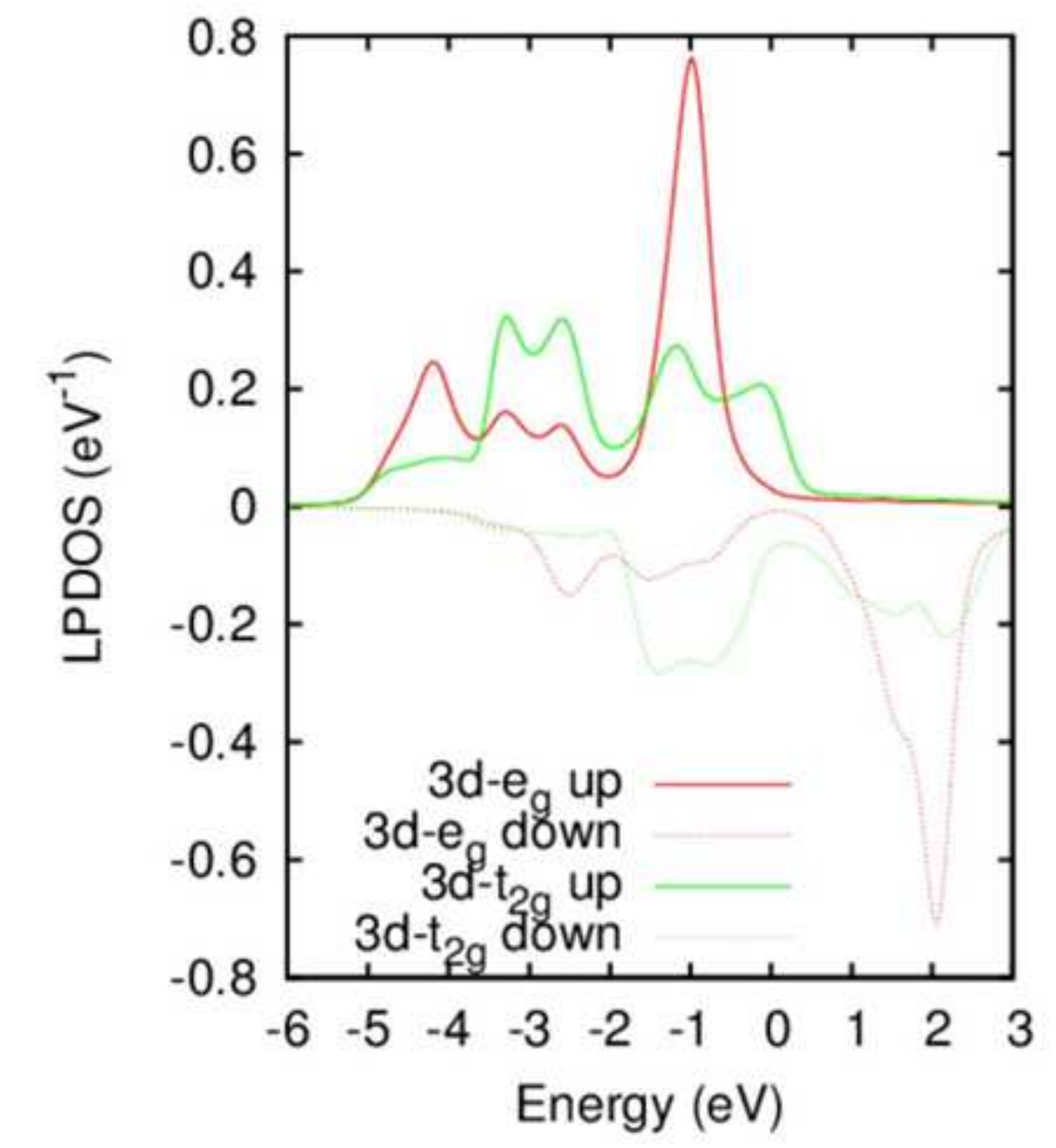}
\caption{\label{fig:pdosfe} Local partial density of states of bulk $\alpha$-Fe. In BCC crystals, the 3$d$ orbitals $xy$, $xz$ and $yz$ are labelled $t_{2g}$ whilst the $x^2-y^2$ and $3z^2-r^2$ are $e_g$.}
\end{figure}

We would like to understand the underlying reason for the relatively poor representation of the magnetic contribution delivered by the Heisenberg Hamiltonian. Fig. \ref{fig:pdos} plots the orbitally projected density of states (PDOS) of the 3$d$-band electrons for the core $\alpha$ and the \nth{1} n.n. $\beta$ atoms for various SIA configurations. The PDOS are calculated using a $6\times 6 \times 6$ MP k-point grid with the same parameters as given in Sec.~\ref{s:method}. A Gaussian broadening scheme with a=0.1eV ($\exp(-(E/a)^2)$) is applied to smooth the PDOS figures. 

For bulk Fe at $a_0^{\text{DFT}}$, the orbitals at the Fermi energy ($E_F$) are predominantly from t$_{\text{2g}}$ states (see Fig\ref{fig:pdosfe}). We would expect that the $t_{\text{2g}}$ and $e_{\text{g}}$ symmetry orbitals contribute differently to magnetic properties. In a recent work by Kvashnin \textit{et. al.} \cite{Kvashnin_prl_2016} and A. Szilva \textit{et. al.} \cite{Szilva_prb_2017}, an orbitally resolved analysis of exchange integrals of BCC Fe revealed that the $t_{\text{2g}}$ orbitals are weakly dependent on the configuration of the spin moments and are \textit{`Heisenberg-like'}, whereas the magnetic behaviour of $e_{\text{g}}$ states originates from double-exchange.

With the exception of the $\langle 111\rangle_D$ configuration, $\beta$ site atoms have a PDOS similar to that in the bulk, while the PDOS of all the $\alpha$ site (core) atoms change significantly. For all the cases, a large van Hove singularity for the majority spin 3$d$ electrons is observed to move into the conduction band. However, in the case of the $\langle 110 \rangle$ dumbbell, the Fermi energy is located at the peak in the PDOS of the 3d$_{xz}$ and 3d$_{xy}$ states, which is 0.5eV lower in energy than the maximum in the t$_{2g}$ PDOS. In the case of a $\langle 100 \rangle$ dumbbell, the 3d$_{z^2}$ and 3d$_{x^2-y^2}$ remain near the Fermi energy in the valence band with the peak in the density of states of $e_g$ orbitals moving into the conduction band. For the minority spin 3$d$ electrons, a deep state at $E_F-4$eV develops, which is not present in the bulk. These deep-state electrons in the Fermi sea will not be easily excited and will not contribute to magnetic excitations, but they give rise to the suppression of magnetic moment at the core of SIA configurations.

The most important distinction in the PDOS calculations is that, unlike the other SIA configurations, the Fermi surface of the $\langle 111 \rangle$ dumbbell is equally characterised by the t$_{\text{2g}}$ and e$_{\text{g}}$ orbitals. $e_{\text{g}}$ orbitals are known not to behave in a Heisenberg-like manner \cite{Kvashnin_prl_2016,Szilva_prb_2017} but will contribute to the low-energy magnetic excitations. Therefore, it is clear why the Heisenberg Hamiltonian is unable to map the $\langle 111\rangle_D$ magnetic contribution well.

\subsection{Self-Consistent Treatment of Longitudinal Fluctuations}\label{ss:landau}

The failure of the Heisenberg Hamiltonian is due to the fact that the energy of formation of a magnetic moment, known also as the band splitting, is not treated by the Heisenberg Hamiltonian. In other words, the itinerant nature of 3$d$-electrons is poorly mapped. A way of improving the description is to include longitudinal magnetic degrees of freedom in the Hamiltonian. A possible way of achieving this is provided by the Heisenberg-Landau Hamiltonian
\begin{equation}\label{eqn:HL}
    \mathcal{H}_{HL} = \mathcal{H}+\displaystyle \sum _i \bigg(A_i\mathbf{S}_i^2 + B_i\mathbf{S}_i^4\bigg),
\end{equation}
where $A$ and $B$ are the Landau coefficients and $\mathcal{H}$ is the Heisenberg Hamiltonian. 

\begin{figure}
\includegraphics[width=8.5cm]{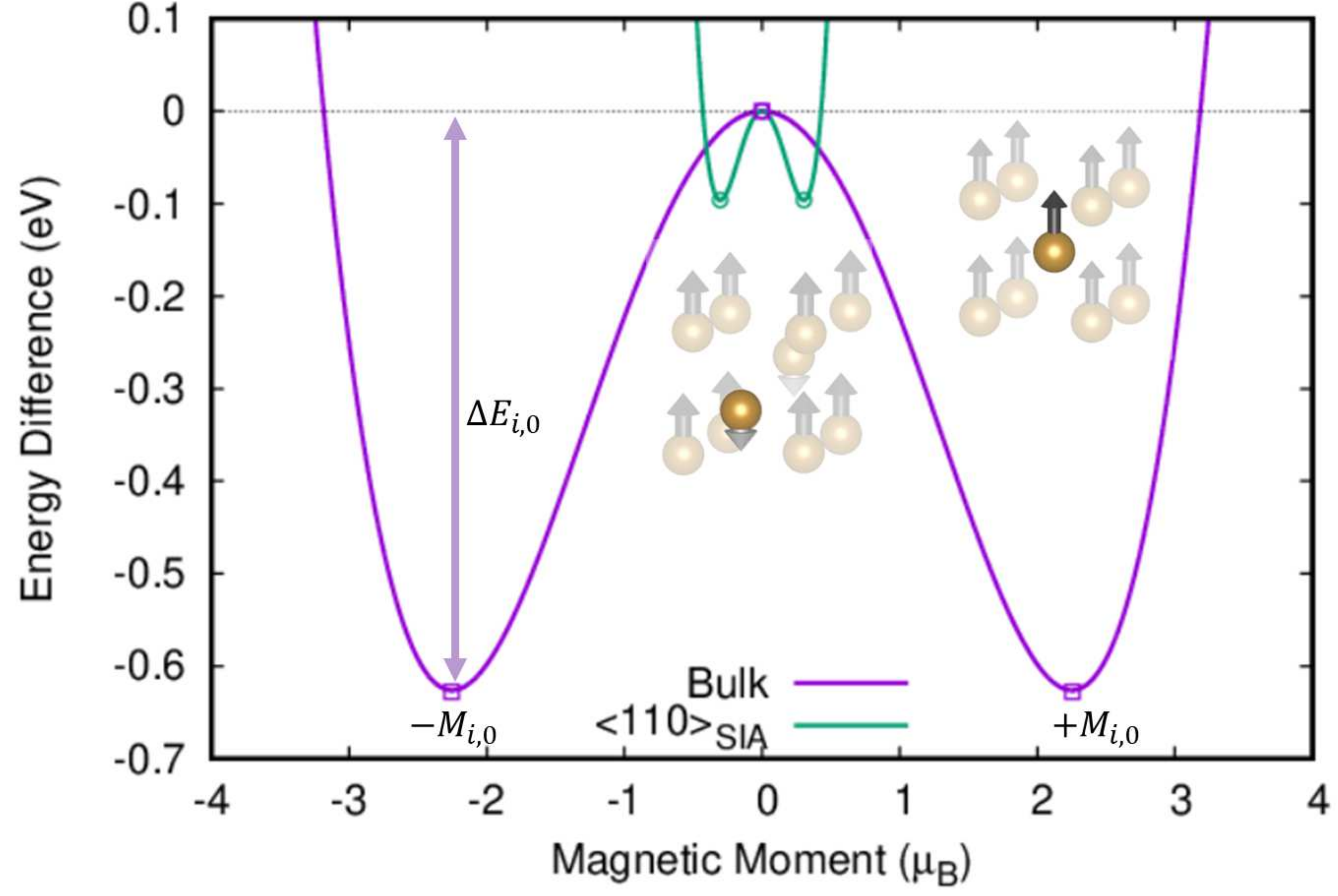}
\caption{\label{fig:landau} The Landau part of energy as a function of the magnitude of magnetic moment. The double well structure is a typical signature of the Landau Hamiltonian. We plotted the energy corresponding to the perfect lattice case and the energy of a core atom in a $\langle 110 \rangle$ dumbbell configuration. The spin-polarized DFT data are shown as points. The curves are drawn using the values of parameters $A_i'=-0.247$ and $B_i'=0.02436$ for a bulk atom, and $A_i'=-2.093$ and $B_i'=11.281$ for the $\langle 110 \rangle$ dumbbell, extracted using Eq. \ref{eqn:deltaE_landau} and \ref{eqn:spinAB}.}
\end{figure}

The Landau terms act to create a double well in the energy with respect to the magnitude of the magnetic moment. The well depth is the energy difference between the magnetic and non-magnetic states. The minimum value is at the spontaneous magnetic moment (Fig.~\ref{fig:landau}). To determine the Landau parameters, we follow the logic presented in Ref. \cite{ma_prb_2017}, where a single set of Landau parameters $A$ and $B$ were constructed for each simulation cell and then parameterised as a function of an effective electron density. Here, we generalise the approach to calculate a set of Landau coefficients for each atomic site.

We begin by defining a Landau Hamiltonian in which we only allow the atomic spin to contribute to the energy of its site $i$, such that we assume no inter-site Landau-type magnetic interactions. The atomic spins act as order parameters:
\begin{equation}\label{eqn:HpL}
    H'_{\text{HL}} = \displaystyle \sum _i \bigg(A'_i\mathbf{S}_i^2 + B'_i\mathbf{S}_i^4\bigg).
\end{equation}

OpenMX allows us to calculate the energy components per site per orbital \cite{openmx,decomp} due to the use of LCPAO (see Appendix \ref{app:decomp}). One can calculate the energy difference for each atomic site between the magnetic and non-magnetic configurations $\Delta E_i$. Since we treat atomic spin as order parameters, the energy difference between any two states for site $i$ can be expressed as
\begin{equation}\label{eqn:deltaE_landau}
    \Delta E_i = A'_i\mathbf{S}_{i,0}^2 + B'_i\mathbf{S}_{i,0}^4
\end{equation}
where $\mathbf{M}_{i,0} = -g\mu_B\mathbf{S}_{i,0} $ is the spontaneous magnetic moment. Knowing that the Landau Hamiltonian should have a minimum at the spontaneous magnetic moment, we are able to derive the moment in terms of the site-resolved Landau coefficients $A'_i$ and $B'_i$:
\begin{equation}
        \frac{\partial \Delta E_i}{\partial \mathbf{S}_{i,0}}=0,
\end{equation}
\begin{equation}\label{eqn:spinAB}
    \implies S_{i,0}=|\mathbf{S}_{i,0}|=\sqrt{\frac{-A'_i}{2B'_i}}\neq 0
\end{equation}
For each atomic site we end up with a pair of simultaneous equations, i.e. Eq. \ref{eqn:deltaE_landau} and \ref{eqn:spinAB}, from which we may determine the site-resolved Landau parameters.

We can then relate the primed Landau coefficients to those in the interacting Heisenberg-Landau Hamiltonian, which also includes the exchange coupling parameters. By equating equations \ref{eqn:HL} and \ref{eqn:HpL}, we find
\begin{align}
    A_i\mathbf{S}_{i}   &= A'_i \mathbf{S}_{i} + \displaystyle \sum _j J_{ij}\mathbf{S}_j\\
                        &= A'_i \mathbf{S}_{i} + \mathbf{h}_i \label{eqn:landau_a}\\
    B_i                 &=B'_i \label{eqn:landau_b}
\end{align}
where $\mathbf{h}_i$ is the effective field of the Heisenberg Hamiltonian, where
\begin{align}
\nonumber
\mathbf{h}_i &= \frac{\partial }{\partial \mathbf{S}_i}\mathcal{H}
= -2\displaystyle \sum_{i\neq j}^{N}J_{ij}\mathbf{S}_{j}.
\label{eqn:ef}
\end{align}

In Table \ref{tab:energy} we evaluate the contribution from the Landau part of the Hamiltonian for each SIA configurations. When the magnetic contribution of the Heisenberg-Landau mapping is added to the non-magnetic DFT energy, the total energy is in excellent agreement with the spin-polarised DFT energy. The small relative error ($\approx 0.1$\%) in the Heisenberg-Landau energies with respect to DFT are well within the inherent error of DFT calculations.

\begin{figure}
\includegraphics[width=8.5cm]{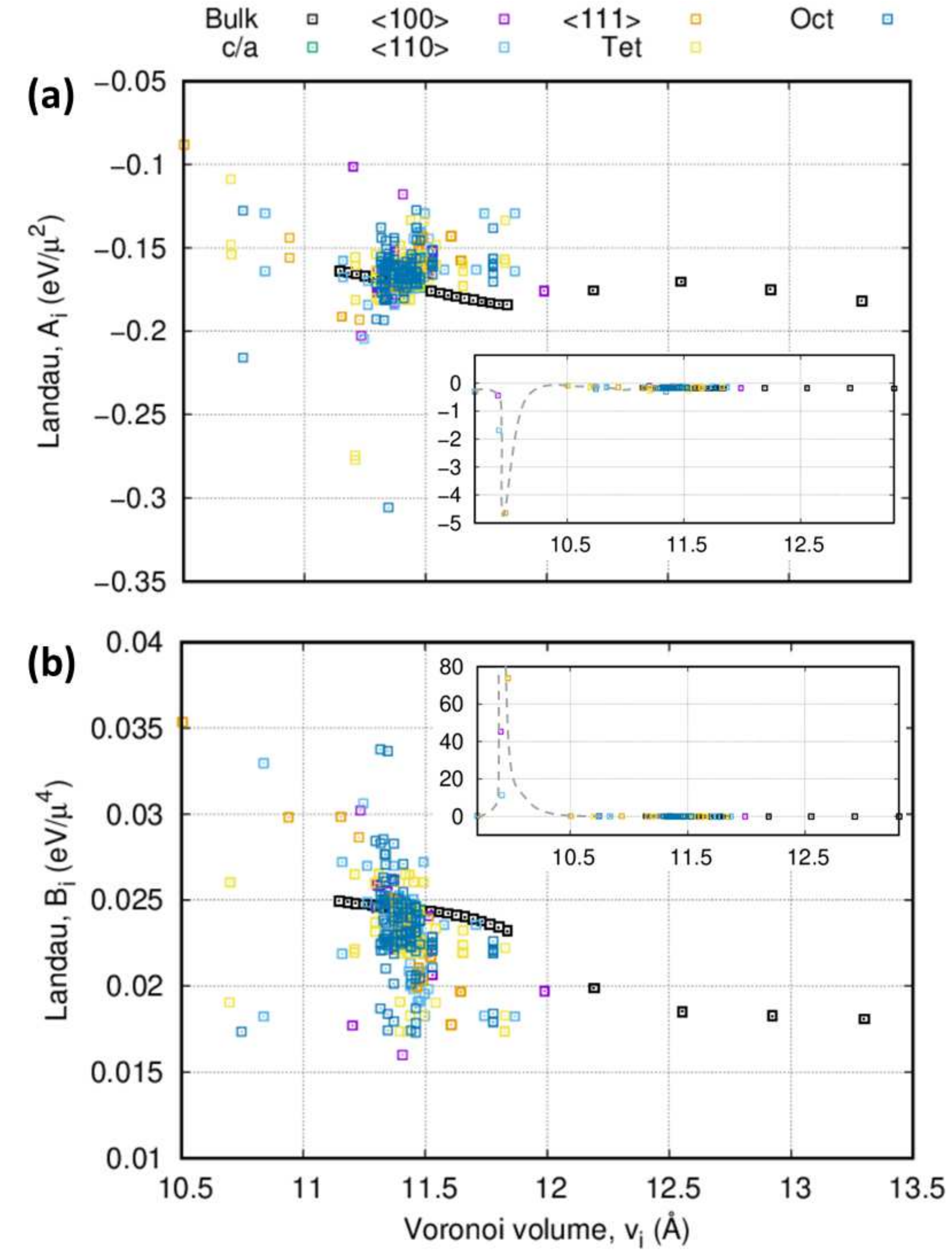}
\caption{\label{fig:landauAB} Values of Landau parameters as a function of the Voronoi volume. Inset figures show the anomalously large values of the Landau terms for core atoms, resulting from the fact that magnetic moments are suppressed. Dashed lines guide the eye and serve no interpretation.}
\end{figure}

In Fig. \ref{fig:landauAB} we plot the resulting Landau coefficients defined according to Eq. \ref{eqn:landau_a} and \ref{eqn:landau_b}, using the values of $J_{ij}$ calculated in Section \ref{ss:sia}. The site resolved coefficients are plotted against the local Voronoi volume calculated using the fuzzy cell partitioning method \cite{Becke_jchemphys_1988}. 


The site resolved mapping confirms that Landau parameters do not form a simple relation with the Voronoi volume. Values of the Landau parameters change significantly for atoms in the core of defects. In Figure~\ref{fig:landauAB_eff} in Appendix~\ref{sec:ef} we also plot the Landau parameters with respect to a tight binding derived effective electron density used in many-body calculations and compare with values derived from a previous bulk definition \cite{ma_prb_2017}. Work is  ongoing on the development of suitable descriptors to accurately represent exchange coupling parameters and Landau parameters with respect to the local environment, which represents a great challenge in the development of an accurate large-scale model combining magnetism and strong lattice deformations. 

\subsection{Low Energy Migration Pathways of Atomic Defects}\label{sec:mig}

\begin{figure*}
\includegraphics[width=16cm]{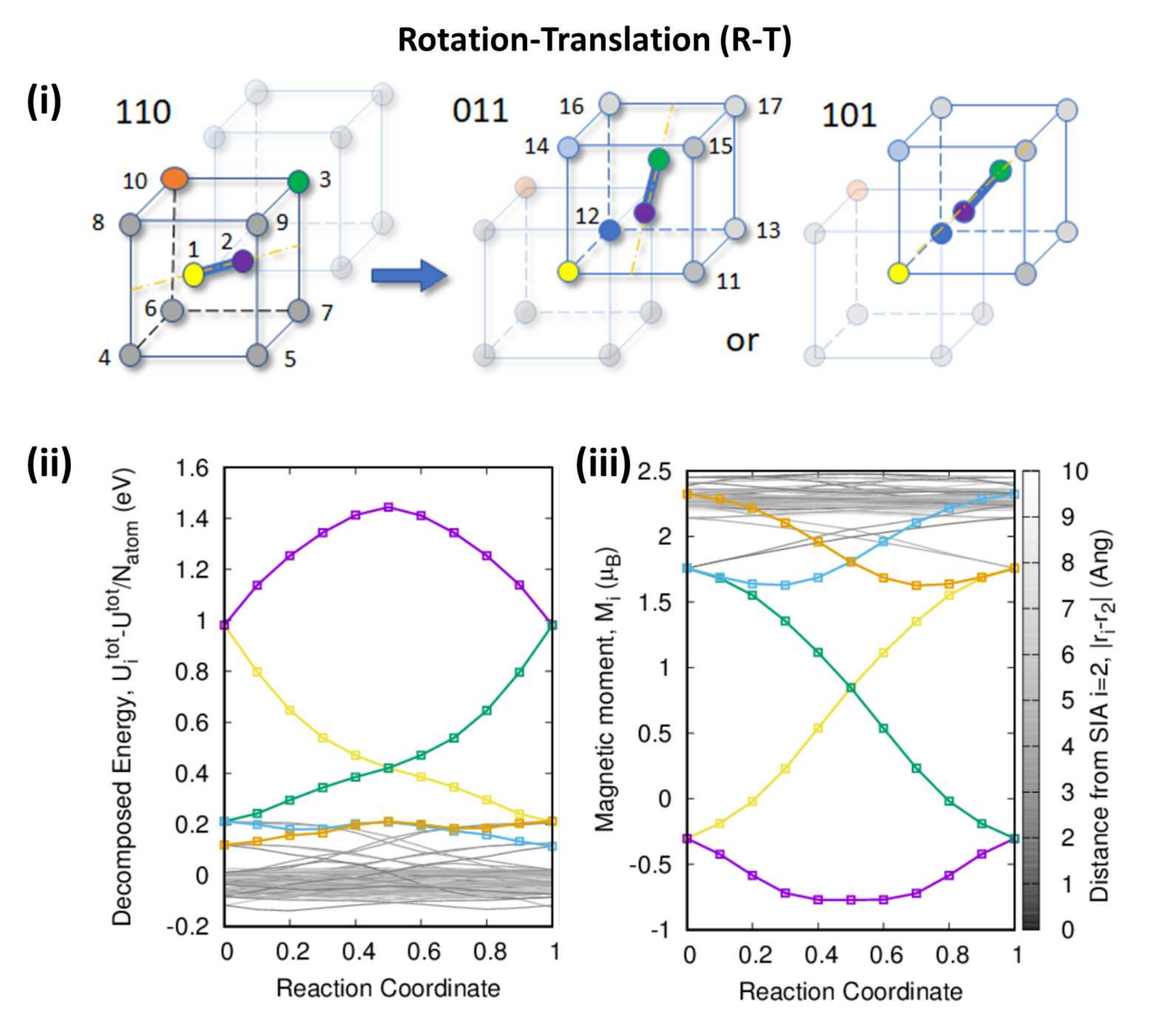}
\caption{\label{fig:emmiga} Migration of $\langle 110 \rangle$ SIA in $\alpha$-Fe via (a) Johnson's mechanism - rotation and translation, and (b) the second neighbour jump - rotation. Part (i) shows a schematic of the transition paths. (ii) and (iii) display the site decomposed energy and magnetic moments, respectively, with selected atoms coloured according to (i). Energies in (ii) are translated relative to the average energy per atom ($U^{tot}=\langle U_i^{tot}\rangle$).}
\end{figure*}

\begin{figure*}
\includegraphics[width=16cm]{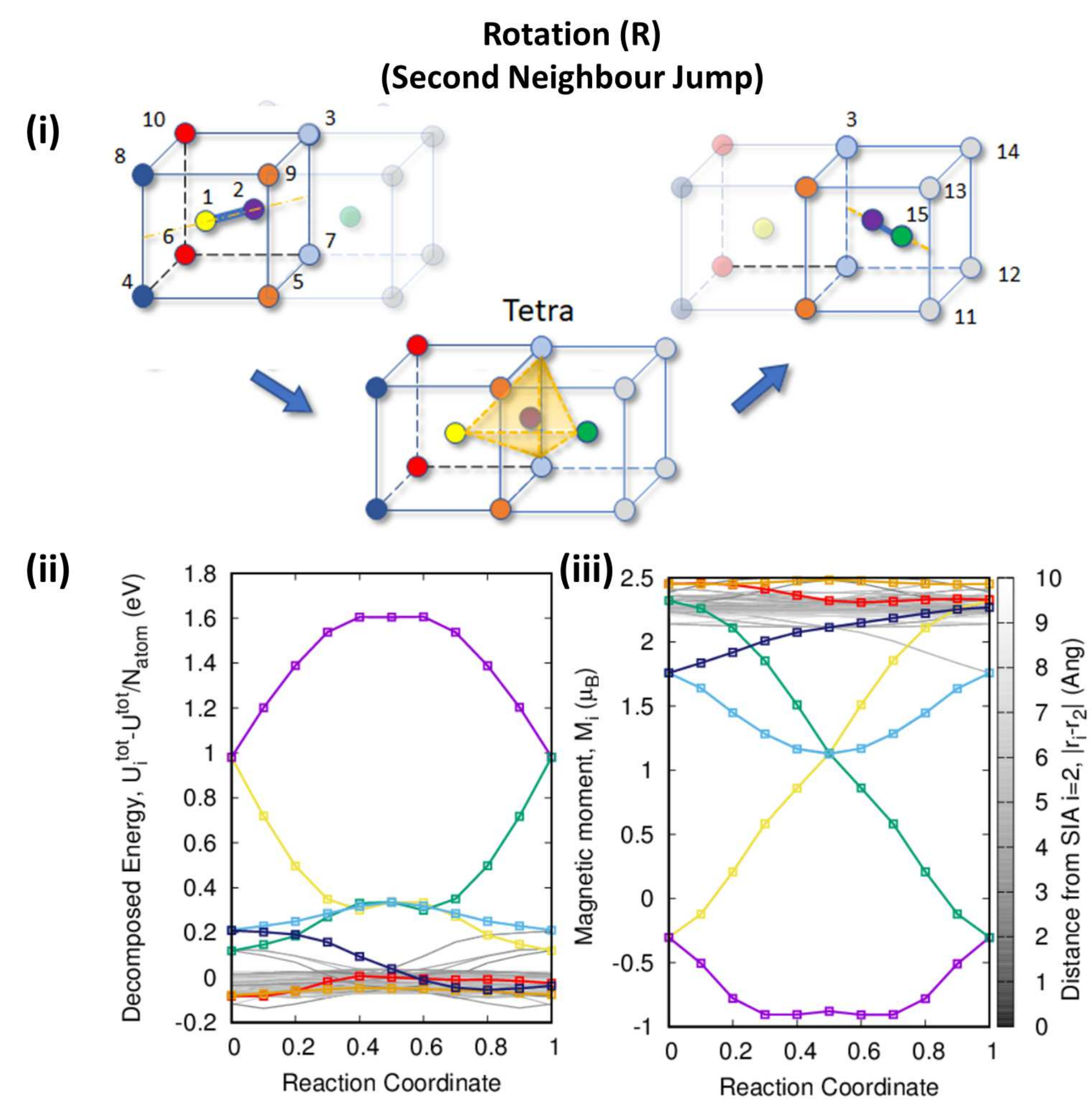}
\caption{\label{fig:emmigb} Migration of $\langle 110 \rangle$ SIA in $\alpha$-Fe via (a) Johnson's mechanism - rotation and translation, and (b) the second neighbour jump - rotation. Part (i) shows a schematic of the transition paths. (ii) and (iii) display the site decomposed energy and magnetic moments, respectively, with selected atoms coloured according to (i). Energies in (ii) are translated relative to the average energy per atom ($U^{tot}=\langle U_i^{tot}\rangle$).}
\end{figure*}

\begin{figure}
\includegraphics[width=8.5cm]{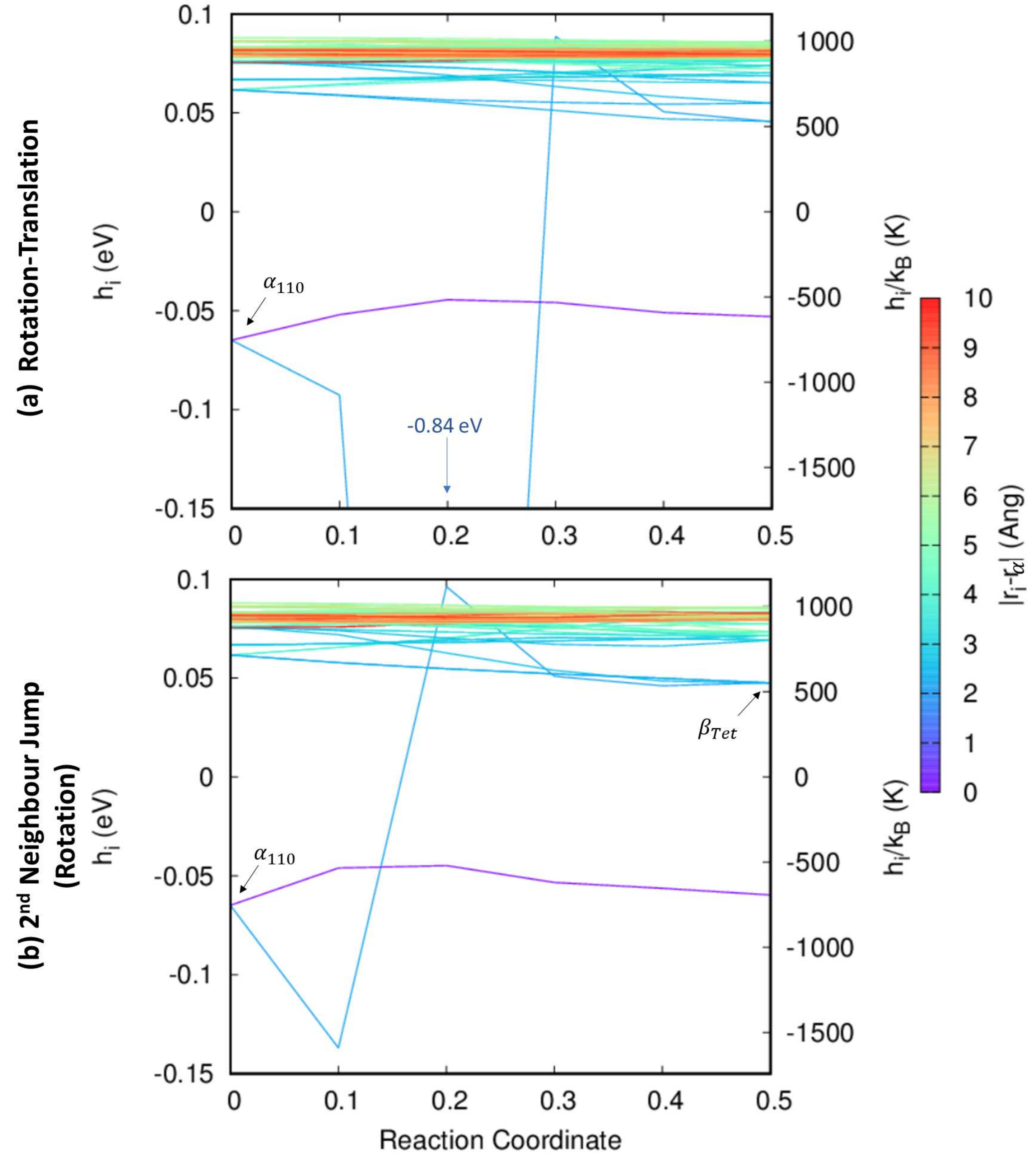}
\caption{\label{fig:efmig}Heisenberg effective field (and temperature) during the migration of a $\langle 110 \rangle_D$ dumbbell in BCC-Fe.}
\end{figure}

In this Section we consider the migration of a $\langle 110 \rangle$ dumbbell via two lowest energy pathways. We calculated the energy barriers and minimum energy pathways using the Nudged Elastic Band method \cite{neb_1,neb_2,neb_3}. Our procedure began with the geometry optimisation of the initial and final configurations as set out in Section \ref{s:method}. Initial energy pathways were constructed by linear interpolation of coordinates between the initial and final configurations.

The lowest barrier for an $\langle 110 \rangle_D$ to overcome is by Johnson's mechanism \cite{johnson_physrev_1964} (simultaneous rotation and translation) with the migration energy 0.34~eV, in excellent quantitative agreement with previous studies  \cite{fu_prl_2004,fu_natmat_2005,Ma_PRM_2019} and experiment \cite{Ehrhart}. The \nth{2} n.n. jump mechanism, via the tetrahedral configuration, has a barrier of 0.49eV. An alternative unfavorable migration path via a purely translational jump via a $\langle 111 \rangle$ configuration has a barrier of 0.79 eV. 

Schematic illustration of mechanisms of rotation-translation migration and rotational migration are shown in Fig. \ref{fig:emmiga} and \ref{fig:emmigb}. In addition, we show the decomposed energy (relative to the average atomic energy) and their respective magnetic moments during the transition, which are used in the calculation of the site resolved Landau parameters. The energy decomposition identifies that the core atoms are approximately 1~eV more energetic than the average.

 
The Heisenberg-Landau Hamiltonian allows us to correctly account for the transverse fluctuations of magnetic moments in the vicinity of the core of SIA configurations. The Landau term adjusts the energy such that the effective magnetic field is zero when a system is in the ground state.  Our DFT calculations are in the adiabatic paradigm such that we expect the effective field on each site to be zero (see the proof of this statement given in Appendix \ref{sec:ef}). We calculate the exchange coupling parameters and Landau coefficients for each NEB image, and verify that such criteria are met.

We may again observe how by only using a Heisenberg Hamiltonian leads to erroneous results. In Fig. \ref{fig:efmig} we show the contribution to the effective field from the Heisenberg term during the rotation-translation and \nth{2} n.n. jumps of a $\langle 110 \rangle$ dumbbell. This interaction may be represented by means of an effective temperature $T_i = |\mathbf{h}_i|/k_B$. Without the Landau terms, core atoms are observed to have a \emph{negative temperature} which is known to exist in nuclear spin systems\cite{purcell_prb_1951}. This occurs as the magnetic moments on the core atoms oppose the effective field due to exchange interactions. However, the negative temperature is merely a consequence of the incompleteness of the Heisenberg model. The Landau terms correct the condition that in the adiabatic regime the effective field acting on each atom must be zero.

\section{Conclusion}


In this study, we explored the connection between  first principles density functional calculations and the use of model Hamiltonians, to describe magnetic interactions in BCC iron containing structural defects. We benchmarked our LCPAO DFT results against literature data to verify the accuracy of calculations performed using the OpenMX code \cite{openmx} and our own in-house exchange coupling code. We are able to correctly explain the known order of stability of self interstitial defects in magnetic Fe: $\langle 110 \rangle\to$ tetrahedral $\to \langle 111 \rangle \to \langle 100 \rangle$, where the comparison of energies derived from magnetic and non-magnetic calculations reveals that magnetism causes the order of stability to change.


We explored the limits of validity of the commonly used Heisenberg Hamiltonian for the description of the magnetic interactions in iron containing defects. Exchange integrals were computed using the magnetic force theorem, allowing us to map the magnetic contribution onto the Heisenberg Hamiltonian functional form. When the mapped Heisenberg magnetic contribution is added to the non-magnetic energies, the energy differences for the self interstitial defects is reproduced within 10\%. The self interstitial configuration where the magnetic energy predicted using the Heisenberg Hamiltonian is poorest is the $\langle 111 \rangle_D$ configuration. This occurs due to the increased population of the $e_g$ orbitals at the Fermi energy.

Failures of the Heisenberg Hamiltonian can be mitigated by adding symmetry-breaking Landau terms to the magnetic Hamiltonian. By projecting energies onto atomic sites we generalise our earlier Landau-Heisenberg Hamiltonian\cite{ma_prb_2017} and define site-resolved Landau coefficients, determining the values of these coefficients directly from the DFT calculations. We show that the Landau terms correct the magnetic energy contribution and provide significantly more accurate representation of energy hypersurfaces, matching magnetic DFT calculations. This information is used for parameterising a new generation of spin-lattice dynamics potentials. We further show how a Heisenberg Hamiltonian can lead to an incorrect interpretation of magnetism in the core of the defects, effectively corresponding to metastable ``negative temperature'' magnetic configurations in the core. These anomalies can be rectified using the Heisenberg-Landau Hamiltonian.



\begin{acknowledgments}
We would like to express our gratitude to Max Boleininger and Andrew London for valuable discussions. This work has been carried out within the framework of the EUROfusion Consortium and has received funding from the Euratom research and training programme 2014-2018 and 2019-2020 under grant agreement No 633053. The views and opinions expressed herein do
not necessarily reflect those of the European Commission.
This work also received funding from the Euratom research and training programme 2019-2020 under grant agreement No. 755039. We acknowledge funding by the RCUK Energy Programme (Grant No. EP/T012250/1), and EUROfusion for providing access to Marconi-Fusion HPC facility in the generation of data used in this manuscript. 
\end{acknowledgments}

\bibliography{references_2019a}

\appendix

\section{Effective Field of Heisenberg-Landau Hamiltonian}\label{sec:ef}

The effective field on an atomic site $k$ in the configuration $\mathcal{R}$ is a measure of the change in energy due to an infinitesimal change in the spin $\mathbf{S}_k$. In our construction of the Heisenberg-Landau Hamiltonian we allow for both transverse and longitudinal fluctuations of the semi-classical spin-vectors:

\begin{align}
    \mathbf{h}^{HL}_k &=\frac{\partial}{\partial \mathbf{S}_k}\mathcal{H}_{HL}(\mathcal{R},\mathcal{S}) \\
    &= -2\displaystyle \sum_i J_{ik}\mathbf{S}_i + 2 A_k \mathbf{S}_k + 4 B_k S_k^2\mathbf{S}_k
\end{align}
When in the electronic ground state, the changes in energy with respect to the spin should at be a minimum in the potential energy surface, thus requiring the effective field (the gradient of this potential) to be zero. This can be easily verified by substituting in the definitions of $A_k$ and $B_k$ from equations \ref{eqn:landau_a}, \ref{eqn:landau_b} and \ref{eqn:deltaE_landau}. In a general spin-state this gives:
\begin{align}
    \mathbf{h}^{HL}_k 
    &= - 2\displaystyle \sum _i J_{ik} \mathbf{S}_i + 4B'_kS_k^2\mathbf{S}_k \nonumber \\
    &\qquad + \frac{2}{\mathbf{S}_{k,0}^2}\bigg( A'_k\mathbf{S}_{k,0}^2 + \displaystyle \sum _i J_{ik}\mathbf{S}_{i,0}\cdot \mathbf{S}_{k,0}\bigg)\mathbf{S}_k\\
    &= - 2\displaystyle \sum _i J_{ik} \mathbf{S}_i + 4B'_kS_k^2\mathbf{S}_k +  2A'_k\mathbf{S}_{k} \nonumber \\
    &\qquad   + 2\displaystyle \sum _i J_{ik}\frac{\mathbf{S}_{k}\cdot\hat{\mathbf{S}}_{k,0}}{S_{k,0}}\mathbf{S}_{i,0} \\
    &= - 2\displaystyle \sum _i J_{ik} \mathbf{S}_i + 4B'_kS_k^2\mathbf{S}_k -  4B'_kS_{k,0}^2\mathbf{S}_{k}  \nonumber \\
    &\qquad  + 2\displaystyle \sum _i J_{ik}\frac{\mathbf{S}_{k}\cdot\hat{\mathbf{S}}_{k,0}}{S_{k,0}}\mathbf{S}_{i,0}\\
    &= 2\displaystyle \sum _i J_{ik} \bigg(\frac{\mathbf{S}_{k}\cdot\hat{\mathbf{S}}_{k,0}}{S_{k,0}}\mathbf{S}_{i,0}-\mathbf{S}_i\bigg) \nonumber \\
    &\qquad + 4B'_k(S_k^2-S_{k,0}^2)\mathbf{S}_k
\end{align}

When the electronic orbitals are in their ground-states for the atomic configuration, then so too will be the spin-order such that $\mathbf{S}_k\xrightarrow[]{GS}\mathbf{S}_{k,0}$:
\begin{align}
    \mathbf{h}^{HL}_k 
    &= 2\displaystyle \sum _i J_{ik} \bigg(\frac{\mathbf{S}_{k,0}\cdot\hat{\mathbf{S}}_{k,0}}{S_{k,0}}\mathbf{S}_{i,0}-\mathbf{S}_{i,0}\bigg) \nonumber \\
    &\qquad + 4B'_k(S_{k,0}^2-S_{k,0}^2)\mathbf{S}_{k,0} \nonumber \\
    &=\mathbf{0}
\end{align}

\begin{figure}
\includegraphics[width=8.5cm]{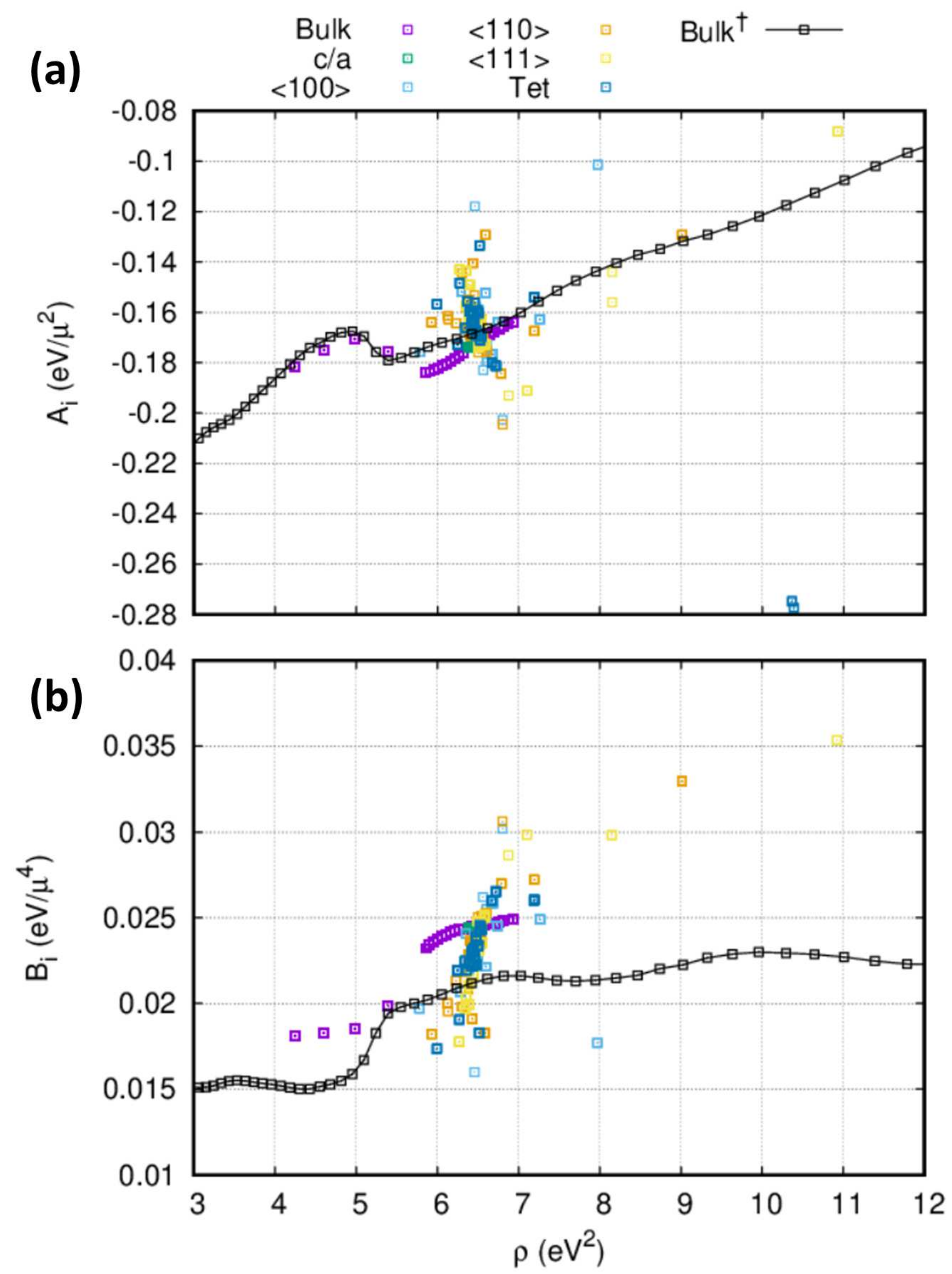}
\caption{\label{fig:landauAB_eff} Extracted Landau parameters as a function of local electron density as defined in Ref~\cite{ma_prb_2017}. The Landau parameters calculated for strained perfect BCC lattice from Ref~\cite{ma_prb_2017} are also shown for comparison (daggered).}
\end{figure}

\section{OpenMX Energy Decomposition}\label{app:decomp}
Due to the finite range of the pseudo-atomic orbitals in the psudopotential based DFT formulation employed within the OpenMX code \cite{openmx,decomp}, the energy can be uniquely decomposed into contributions from each atomic site ($i$) and localised orbital ($\alpha$):
\begin{align}
    E_{Tot} &= E_{kin} + E_{ec} + E_{ee} + E_{xc} + E_{cc}\\
            &= E_{kin} + (E_{ec}^{(L)} + E_{ec}^{(NL)}) + E_{ee} + E_{xc} + E_{cc}
\end{align}
where the total energy terms are the kinetic energy ($E_{kin}$), electron-core Coulomb energy ( $E_{ec}$), electron-electron Coulomb energy ($E_{ee}$), exchange correlation energy ($E_{xc}$) and the core-core Coulomb energy ($E_{cc}$). For practical and efficient implementation, OpenMX reorganises the electron and core terms into two short range, and one long range term:
\begin{equation}\label{eqn:etot}
    E_{Tot} = E_{kin} + E_{xc} + E_{na} + E_{ec}^{(NL)} + E_{\delta ee} + E_{scc}.
\end{equation}
Each term can be reduced into contributions from site and orbital indices:
\begin{equation}
    E_{Tot} = \displaystyle \sum _{i\alpha} E_{i\alpha}
\end{equation}

%
%
The kinetic energy operator can be decomposed as:

\begin{align}
E_{kin} &= \displaystyle \sum _{\sigma} \displaystyle \sum_{i\alpha}\bigg( \displaystyle \sum_{j\beta} \displaystyle \sum _n^N \rho^{(\mathbf{R_n})}_{\sigma,i\alpha,j\beta}h^{(\mathbf{R_n})}_{i\alpha,j\beta,kin}\bigg)\\
        &= \displaystyle \sum _{\sigma} \displaystyle \sum_{i\alpha} E_{\sigma,i\alpha,kin}
\end{align}
where the matrix elements of the kinetic energy operator are defined as:
\begin{equation}
    h^{(\mathbf{R_n})}_{i\alpha,j\beta,kin} = \frac{1}{V_B}\int_{BZ}dk^3 \displaystyle \sum _{\mu}^{Occ} \langle \psi ^{(\mathbf{k})}_{\sigma \mu} | \hat{T} |  \psi ^{(\mathbf{k})}_{\sigma \mu} \rangle
\end{equation}

%
%
The electron-core Coulomb terms:
\begin{align}
    E^{(NL)}_{ec}  &= \displaystyle \sum _{\sigma} \displaystyle \sum _{i\alpha} \bigg(\displaystyle \sum _n ^N  \displaystyle \sum _{j\beta} \rho^{(\mathbf{R_n})}_{\sigma i \alpha j \beta} h^{(\mathbf{R_n})}_{i\alpha,j\beta,NL} \bigg)\\ 
    &= \displaystyle \sum _{\sigma} \displaystyle \sum _{i\alpha} \bigg(\displaystyle \sum _n ^N  \displaystyle \sum _{j\beta} \rho^{(\mathbf{R_n})}_{\sigma i \alpha j \beta} \langle \phi_{i\alpha} | \displaystyle \sum _{I} V_{NL,I} | \phi _{j\beta} \rangle \bigg) \\
   &= \displaystyle \sum _{\sigma} \displaystyle \sum _{i\alpha} E^{(NL)}_{\sigma i \alpha}
\end{align}
where $V_{NL}$ is the non-local part of the pseudopotential.

%
%
The neutral atom term:
\begin{align}
    E_{na}  &= \int dr n(r)V_{na,I} \\
            &= \displaystyle \sum_{\sigma} \displaystyle \sum_{i\alpha} \bigg( \displaystyle \sum _{j\beta} \displaystyle \sum _n^N \rho^{(\mathbf{R_n})}_{\sigma, i\alpha, j\beta} h^{(\mathbf{Rn})}_{i\alpha,j\beta,na} \bigg)\\
            &=  \displaystyle \sum_{\sigma} \displaystyle \sum_{i\alpha} E^{na}_{\sigma,i\alpha}
\end{align}

%
%
Screened core correction:
\begin{align}
    E_{scc} &= \frac{1}{2} \displaystyle \sum_{I,J} \bigg(\frac{Z_I Z_J}{|\tau_I-\tau_J|} - \int dr n^{(a)}_I(r)V^{(a)}_{H,J}(r)\bigg) \\
            &= \displaystyle \sum _{\sigma} \displaystyle \sum_{i\alpha} \bigg( \frac{1}{2N_i}\displaystyle \sum_j \frac{Z_I Z_J}{|\tau_I-\tau_J|} \nonumber \\
            &\quad - \int dr n^{(a)}_I(r)V^{(a)}_{H,J}(r)\bigg) \\
            &= \displaystyle \sum _{\sigma} \displaystyle \sum_{i\alpha} E^{scc}_{\sigma,i\alpha}
\end{align}

%
%
Electron electron Coulomb term:
\begin{align}
    E_{\delta ee} &= \frac{1}{2} \int dr \big( n(r) - \displaystyle \sum_I n_I^{(a)}(r)\big)\delta V_H(r) \\
        &= \displaystyle \sum_{\sigma} \displaystyle\sum_{i\alpha} \frac{1}{2}\bigg( \displaystyle\sum _{j\beta} \displaystyle\sum_n^N \rho^{(\mathbf{R_n})}_{\sigma,i\alpha,j\beta} h^{\delta V} _{i\alpha,j\beta} \nonumber \\
        &\quad - \frac{1}{2} \int dr \frac{n^{(a)}_i(r)}{N_{i}}\delta V_H(r) \bigg) \\
        &= \displaystyle \sum_{\sigma} \displaystyle\sum_{i\alpha} E^{\delta ee}_{\sigma,i\alpha}
\end{align}

%
%
The exchange correlation term:
\begin{align}
    E_{xc}  &=  \int dr \big( n(r) + n_{pcc}(r) \big) \epsilon_{xc}(r) \\
            &= \displaystyle \sum _{\sigma} \displaystyle \sum _{i\alpha} \bigg( \displaystyle \sum_n^N \displaystyle\sum_{j\beta} \rho^{(\mathbf{R_n})}_{\sigma,i\alpha,j\beta,xc} h^{(\mathbf{R_n})}_{i\alpha,j\beta,xc} \\
            & + \frac{1}{2N_i}\int dr n_{pcc,i}(r)\epsilon_{xc}(r) \bigg) \\
            &= \displaystyle \sum _{\sigma} \displaystyle \sum _{i\alpha} E^{xc}_{\sigma,i\alpha}
\end{align}

The proof of the decomposition for each term in the KS-Hamiltonian (eqn.~\ref{eqn:etot}) can be found in the Openmx developer material by T. Ozaki \cite{decomp}:  http://www.openmx-square.org/workshop/meeting15/.
For the completeness, we just copy their note here. Our reworked derivation is available upon request from the CCFE Publications Manager.

Projections of non-local terms in the energy are treated as mean-field
We confirm in this method the sum of the atomic resolved energies is equivalent to the total energy in the Kohn-Sham DFT calculation (eqn 7). 

\end{document}